%% file: main.tex
\documentclass[sigconf]{acmart}

\AtBeginDocument{%
  \providecommand\BibTeX{{%
    \normalfont B\kern-0.5em{\scshape i\kern-0.25em b}\kern-0.8em\TeX}}}
\copyrightyear{2021}
\acmYear{2021}
\setcopyright{acmcopyright}
\acmConference[ACSAC '21]{Annual Computer Security 
Applications Conference}{December 6--10, 2021}{Virtual Event, USA}
\acmBooktitle{Annual Computer Security Applications Conference (ACSAC '21), 
December 6--10, 2021, Virtual Event, USA}
\acmPrice{15.00}
\acmDOI{10.1145/3485832.3485917}
\acmISBN{978-1-4503-8579-4/21/12}

\usepackage{balance}
\usepackage[utf8]{inputenc}
\usepackage{graphicx}
\usepackage{url}
\usepackage[ruled,vlined]{algorithm2e}

\usepackage{amssymb}

\newcommand{\ignore}[1]{}

\newcommand{\cmark}{\checkmark}
\newcommand{\xmark}{\textit{\sffamily X}}

\begin{document}
\fancyhf{}
\title{SMap: Internet-wide Scanning for Spoofing}

\author{Tianxiang Dai}
\affiliation[obeypunctuation=true]{%
  \institution{ATHENE Center},
  \country{Germany}
}
\affiliation[obeypunctuation=true]{%
  \institution{Fraunhofer SIT},
  \country{Germany}
}

%\affiliation[obeypunctuation=true]{%
%  \institution{Fraunhofer Institute for Secure Information Technology SIT},
%  \country{Germany}
%}
%\affiliation[obeypunctuation=true]{%
%  \institution{National Research Center for Applied Cybersecurity ATHENE},
%  \country{Germany}
%}

\author{Haya Shulman}
\affiliation[obeypunctuation=true]{%
  \institution{ATHENE Center},
  \country{Germany}
}
\affiliation[obeypunctuation=true]{%
  \institution{Fraunhofer SIT},
  \country{Germany}
}

\input{abstract.tex}



 %% TODO: update keywords

\maketitle

\input{introduction.tex}
\input{related-work}
\input{tools}
\input{measurements.tex}
\input{net-analysis.tex}
\input{conclusions.tex}

%\section*{Acknowledgements}
\begin{acks}
This work has been co-funded by the German Federal Ministry of Education and Research and the Hessen State Ministry for Higher Education, Research and Arts within their joint support of the National Research Center for Applied Cybersecurity ATHENE and by the Deutsche Forschungsgemeinschaft (DFG, German Research Foundation) SFB~1119.
\end{acks}

%\newpage
\bibliographystyle{ACM-Reference-Format}
\bibliography{bib,bib2,rfc}
\balance

%\newpage
%\appendix
%\input{false.tex}

\end{document}

%% file: abstract.tex
\begin{abstract}
To protect themselves from attacks, networks need to enforce {\em ingress filtering}, i.e., block inbound packets sent from spoofed IP addresses. Although this is a widely known best practice, it is still not clear how many networks do not block spoofed packets. Inferring the extent of spoofability at Internet scale is challenging and despite multiple efforts the existing studies currently cover only a limited set of the Internet networks: they can either measure networks that operate servers with faulty network-stack implementations, or require installation of the measurement software on volunteer networks, or assume specific properties, like traceroute loops. Improving coverage of the spoofing measurements is critical.

In this work we present the {\bf Spoofing Mapper (SMap)}: the first scanner for performing {\em Internet-wide} studies of ingress filtering. SMap evaluates spoofability of networks utilising standard protocols that are present in almost any Internet network. We applied SMap for Internet-wide measurements of ingress filtering: we found that 69.8\% of all the Autonomous Systems (ASes) in the Internet do not filter spoofed packets and found 46880 new spoofable ASes which were not identified in prior studies. Our measurements with SMap provide the first comprehensive view of ingress filtering deployment in the Internet as well as remediation in filtering spoofed packets over a period of two years until May 2021. %SMap is simple to use and does not require installation on the tested network nor coordination with the tested networks. 

%, in contrast to 2.4\% ASes identified by the most recent study of the Spoofer Project, as well as 46880 new spoofable ASes.
%We developed and integrated into SMap three techniques for measuring ingress filtering (based on IPID, DNS, PMTUD) and applied them 

%Nevertheless, modular architecture of SMap allows integrating new measurement techniques based on other protocols.

%SMamp also provide a much larger coverage, 30\% more ASes, than the previous projects.
%Our tools can be applied for periodic Internet wide measurement of ingress filtering and provide 
%as well as notification of networks which allow packets from spoofed IP addresses into their networks.

We set up a web service at \url{https://smap.cad.sit.fraunhofer.de} to perform continual Internet-wide data collection with SMap and display statistics from spoofing evaluation. We make our datasets as well as the SMap (implementation and the source code) publicly available to enable researchers to reproduce and validate our results, as well as to continually keep track of changes in filtering spoofed packets in the Internet.
\end{abstract}
% evaluations and enables continual Internet-wide data collection, its analysis and reporting of the statistics to the web page. We also
% We make our datasets available.
%the existing studies do not allow to infer the extent at Internet-scale of the networks that enforce ingress filtering, providing results with only a limited coverage of the Internet networks

%% file: introduction.tex
\section{Introduction}
% Ingress filtering can cause problems with multihoming.

Source IP address spoofing allows attackers to generate and send packets with a false source IP address impersonating other Internet hosts, e.g., to avoid detection and filtering of attack sources, to reflect traffic during Distributed Denial of Service (DDoS) attacks, to launch DNS cache poisoning, for spoofed management access to networking equipment and even to trigger services which can only be accessible to internal users \cite{rossow2014amplification,brandt2018domain,qian:ccs20,letsencrypt,dai2021hijackers}. The best way to prevent IP spoofing is by enforcing Source Address Validation (SAV) on packets, a practice standardised in 2000 as BCP38 \cite{rfc2827}: {\em ingress filtering} for blocking inbound packets and {\em egress filtering} for blocking outbound packets sent from spoofed IP source addresses. %Ingress filtering appears to be an easier problem to solve in practice: the network need to only filter according to the list of the , with only a couple of recent measurement studies proving information on the fraction of networks without ingress filtering. 
In contrast to egress filtering which has been extensively measured in the last 15 years, only a couple of recent studies provided measurements on the extent of ingress filtering.

% Why is ingress filtering a more easier problem to solve?

{\bf Ingress filtering.} To enforce ingress filtering the networks should check the source address of an inbound packet against a set of permitted addresses before letting it into the network. Otherwise, the attackers using spoofed IP addresses belonging to the network can trigger and exploit internal services and launch attacks. For instance, by spoofing internal source IP addresses the attackers can obtain access to services, such as RPC, or spoofed management access to networking equipment [RFC3704], the attackers can cause DoS amplification by triggering the ICMP error messages from the attacked hosts to other internal hosts whose IP addresses the attacker spoofed. Enforcing ingress filtering is therefore critical for protecting the networks and the internal hosts against attacks. Nevertheless, despite efforts to prevent IP spoofing, it is still a significant problem. Attacks utilising IP spoofing remain widespread \cite{ferguson2000network,specht2004distributed,miao2015dark,czyz2014taming,brandt2018domain,crime-research}.
%- this prevents attacks on internal hosts from external attackers. 

%To prevent attacks on external hosts, networks should also perform egress filtering on outbound traffic, i.e., block packets sent to the Internet from the network with source IP address not in the network.
%For instance, consider the echo UDP port 7 for such an exploit. UDP port 7 is typically the echo service provided to hosts on the network, e.g., for trouble shooting purposes and performance monitoring. The service echos whatever data is sent to it back to the source. An attack exploiting this is to send a packet with source IP address of one host on the network, from port 7 to another host to port 7. This would create an infinite loop, whereby each of the two hosts would be echoing packets to one another. Another example of this attack is a ``fraggle'' attack, originated with a broadcast message exploiting echo and chargen UDP services, \cite{specht2004distributed}. There are many other UDP services that are vulnerable to such exploits and also TCP services that can be exploited, see list in \cite{crime-research}. To prevent exploits of these services, the networks should perform ingress IP filtering, i.e., block packets from the Internet with source IP address in the network.

%How many network deploy ingress filtering to block attacks from the outside on internal hosts? 
{\bf How widespread is the ability to spoof?} There are significant research and operational efforts to understand the extent and the scope of (ingress and egress)-filtering enforcement and to characterise the networks which do not filter spoofed packets; we discuss these in Related Work, Section \ref{sc:works}. Although the existing studies and tools, such as the Open Resolver \cite{mauch2013open} and the Spoofer \cite{beverly2005spoofer, beverly2009understanding, beverly2013initial, lone2018using, luckie2019network} projects, provide a valuable contribution for inferring networks which do not enforce spoofing, they are nevertheless insufficient: they provide a meager (often non-uniform) coverage of the Internet networks and are limited in their applicability as well as effectiveness. %; details follow in Section \ref{sc:limitations}.

{\bf SMap (The Spoofing Mapper).} In this work we present the first Internet-wide scanner for networks that filter spoofed inbound packets, we call the Spoofing Mapper (SMap). We apply SMap for scanning ingress-filtering in more than 90\% of the Autonomous Systems (ASes) in the Internet. The measurements with SMap show that more than 80\% of the tested ASes do not enforce ingress filtering (i.e., 72.4\% of all the ASes in the routing system), in contrast to 2.4\% identified by the latest measurement of the Spoofer Project \cite{luckie2019network}. The reason for this significant difference is the limitation of the previous studies of ingress filtering to a small set of networks. 
% to check the percentage of ASes scanned by previous studies.
%30 ASes discovered in Spoofer were not covered by us (maybe they opted out from sonar project). They are not covered in our technique because they opted out Sonar project and hence not included in our dataset.

%How many ASes in the Internet did you test: 61000 ASes (Sonar dataset). In Sonar you receive IP. Entire IPv4 scan.
% Geolight GeoIP DB: other DB. Affected 
% 90 

%SMap provides means to track deployment of ingress filtering in the Internet.

%\label{sc:limitations}
{\bf Limitations of filtering studies.} The measurement community provided indispensable studies for assessing ``spoofability'' in the Internet, and has had success in detecting the ability to spoof in some individual networks using active measurements, e.g., via agents installed on those networks \cite{mauch2013open,lone2018using}, or by identifying spoofed packets using offline analysis of traffic, e.g., \cite{lone2017using,luckie2019network}. The need to install agents on networks or the ability to obtain traces only from some networks limits the studies to non-uniform coverage of the Internet. Therefore it is not clear how representative these statistics are.
Unfortunately, this limitation to a small set of networks creates a bias in the assessments of the overall number of spoofable networks. The extrapolation from the small set of networks to the entire Internet typically result in assessment that at least 30\% of the Internet networks do not filter spoofed packets \cite{luckie2019network,qian:ccs20}. As we show, the number of spoofable networks is above 72\% which is significantly higher than what was previous believed.

{\bf Requirements on Internet studies.} The key requirements for conducting Internet studies upon which conclusions can be drawn include scalable measurement infrastructure, good coverage of the Internet and a representative selection of measurement's vantage points. We summarise the limitations of the previous studies below and in Table \ref{tab:related:work}, and compare to SMap.

$\bullet$ {\em Limited coverage.} Previous studies infer spoofability based on measurements of a limited set of networks, e.g., those that operate servers with faulty network stack \cite{kuhrer2014exit} or networks with volunteers that execute the measurement software \cite{beverly2005spoofer, beverly2009understanding, mauch2013open,beverly2013initial, lone2018using, luckie2019network}, or networks that agree to cooperate and volunteer their traffic logs for offline analysis, e.g., \cite{luckie2019network}. In contrast, the measurements with SMap use standard protocols supported by almost any network with Internet connectivity, for the first time providing studies of ingress filtering that cover the entire IPv4 space. 

% (offline analysis misses out on the networks which do not enforce filtering but which did not sent/received packets from spoofed IP addresses in that time frame).
%(at least during the time frame in which the traces analysis was performed), and hence does not provide an indication on the deployment of ingress filtering in the Internet. 
$\bullet$ {\em Limited scalability.} Previous approaches require installing agents, need to reproduce loops in traceroutes, or use misconfigurations in networks which limits their scalability. SMap is more scalable than any previous approach, since it merely exchanges requests/responses with networks using a fixed infrastructure of probers. The measurement infrastructure of SMap is not a function of the measured networks, hence adding more networks to the study does not require extending the measurement infrastructure.

$\bullet$ {\em Limited representativeness.}  Volunteer or crowd-sourcing studies, such as the Spoofer Project \cite{lone2018using}, are inherently limited due to bias introduced by the participants. These measurements are performed using a limited number of vantage points, which are set up in specific networks, and hence are often not representative of the entire Internet. Increasing the coverage and selecting the networks more uniformly is imperative for collecting representative data; \cite{huz2015experience} showed that the measured network significantly influences the resulting data as well as the derived conclusions. Since SMap measures almost all the IPv4 networks the results are representative of the entire Internet.

%or an attachment point from which the data is collected 
% Small coverage limits the robustness of extrapolation.
% \cite{beverly2013initial}

$\bullet$ {\em Limited stability.} Current measurement studies use unstable infrastructures: volunteers running agents can reinstall computers or move to other networks \cite{mauch2013open}; misconfigured servers \cite{lone2018using} (e.g., with open resolution or with faulty network stack) can be updated -- all causing the network to ``disappear from the radar'' although it may still be spoofable. Hence, longitudinal studies, such as the Spoofer Project, are biased by the stability of the vantage points, and cannot accurately track deployment of ingress filtering in individual networks. A few works \cite{mauch2013open} pointed out that the instability of the infrastructure creates discrepancy in the statistics. In particular, repeating the measurements a few weeks later generates other different results.  

\begin{figure}
\centering
\includegraphics[width=0.45\textwidth]{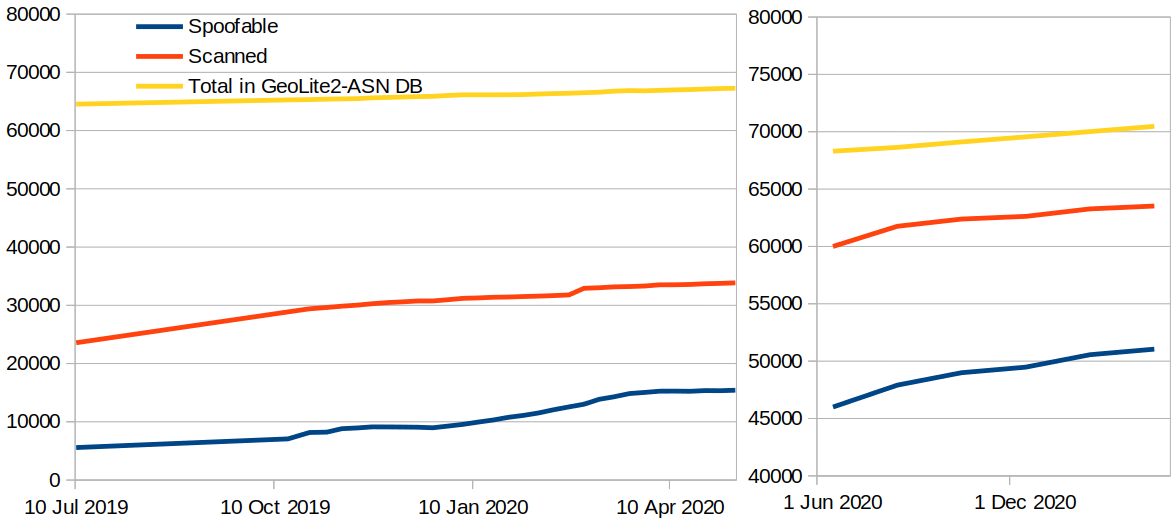}
\caption{SMap measurements between July'19 and May'21. Domain-based (left) and IPv4-based (right).}
\label{fig:measure-ases-timeseries}
%\vspace{-10pt}
\end{figure}

{\bf What SMap improves.} The infrastructure of SMap is more stable than those used in previous studies, e.g., we do not risk volunteers moving to other networks. Our measurements do not rely on misconfigurations in services which can be patched, blocking the measurements. The higher stability also allows for more accurate reproduction and validation of our datasets and results, and enables to perform reliable longitudinal studies. We ran ingress filtering measurements with SMap every week over a period of two years (between 10 July 2019 and 10 May 2021). Our results plotted in Figure \ref{fig:measure-ases-timeseries} demonstrate that the number of spoofable ASes is stable and proportionally increases with the growth in the overall number of ASes in the Internet. This is in contrast to previous studies, e.g., \cite{lone2017using,lichtblau2017detection,lone2018using}, in which a repeated evaluation even a week later provided different statistics. Our two year long measurements between 2019 and 2021 of more than 90\% of Internet's ASes we found 50,023 new ASes that do not enforce ingress filtering, which were not known before, and confirmed all the other ASes that were found spoofable in prior studies.
%Ingress filtering measurements with SMap, plotted in Figure \ref{fig:measure-ases-timeseries}, which we performed between 10 July 2019 and 10 May 2021, demonstrate that the number of spoofable ASes is stable and proportionally increases with the growth in the overall number of ASes in the Internet.

{\bf Ethical Considerations.} Internet-wide scans are important for security research \cite{lyon2009nmap,durumeric2013zmap} and have proven valuable in improving the security landscape of the Internet, including exposing new vulnerabilities, tracking adoption of defences. Nevertheless, Internet-wide scans introduce also ethical challenges. We communicated with network operators to understand and consider the ethical implications of Internet-wide scans. We identified two issues as particularly important for our measurements: {\em traffic load} and {\em consent}.

$\bullet$ {\em Traffic load.} Network scans, such as \cite{lyon2009nmap,durumeric2013zmap,kuhrer2014exit}, require exchanging packets with a large number of Internet networks as well as IP addresses inside the networks. To avoid scanning the Internet we periodically download a dataset of a full scan of the Internet done by Sonar.

%services that are needed for our scan from We develop a series of measures in design of SMap to reduce the traffic load of our scans on the Internet and on the measured networks. In addition to supporting the mere scan of the IPv4 address range, we also integrated support for domain-based scans - this allows testing networks for ingress filtering without scanning them, hence remarkably reducing the traffic volume. We explain this in Section \ref{sc:tools}. We further integrated large inter-scan delays to reduce the load by spreading the measurements traffic over longer time periods.

 %When scanning domains we do not need to search for open ports on every IP address, but instead query the domain for services directly. Nevertheless, IPv4 address range is left as an option in our implementation to enable the research or operational community to use it if they decide that the benefits of SMap scans overwhelm the load that it would introduce. 

$\bullet$ {\em Consent of the scanned.} It is often impossible to request permission from owners of all the tested networks in advance, this challenge similarly applies to other Internet-wide studies \cite{lyon2009nmap,durumeric2013zmap,durumeric2014matter,kuhrer2014exit}. Like the other studies, \cite{durumeric2013zmap,durumeric2014matter}, we provide an option to opt out of our scans. To opt out the network has to provide either its network block (in CIDR notation), domain or ASN through the contact page at \url{https://smap.cad.sit.fraunhofer.de}. Performing security scans is important - the networks that do not enforce filtering of spoofed packets pose a hazard not only to their operators but also to their users, customers and services, as well as other networks. Due to the importance of identifying such networks, in their recent study \cite{luckie2019network} even make public the (``name-and-shame'') lists of providers with missing or misconfigured filtering of spoofed packets; \cite{luckie2019network} also discuss stronger measures against spoofable networks, including liability for damages, and various types of regulation. Inevitably, due to the risks that such networks pose to the Internet ecosystem, it is of public interest to know who those networks are. We do not make the identity of the networks, that do not filter spoofed packets, publicly available, but inform the general public on the fraction of such networks and provide their characterisation (i.e., size, geo-location, business type) in Section \ref{sc:nets}.

Undoubtedly, filtering spoofed packets is critical and networks have to deploy best practices, such as BCP38 \cite{rfc2827} and BCP84 \cite{rfc3704}, to ensure security of the Internet ecosystem. Understanding the extent of filtering is also significant for devising future policies, defence mechanisms or estimating threats and risks to attacks.

{\bf Organisation.} Our work is organised as follows: we compare our study and SMap to related work in Section \ref{sc:works}. In Section \ref{sc:tools} we present the design and the implementation of SMap and the measurement techniques that it uses. In Section \ref{sc:data} we report on the data collected with SMap and the statistics that we derived from it. We characterise the networks which we found not to enforce ingress filtering in Section \ref{sc:nets}. We conclude this work in Section \ref{sc:conc}.

%In Denial of service (DoS) attacks the attackers ustilise IP spoofing to make filtering of the attack more difficult and to avoid exposure.
%Ingress filtering was long proposed as the most effective mechanism for defeating DoS attacks which deploy IP spoofing \cite{ferguson2000network}.

%% file: related-work.tex
\section{Overview of Spoofing Studies}\label{sc:works}
%- representativeness of the studied networks. How representative the data is of the larger Internet? \cite{huz2015experience}.

\subsection{Egress vs. Ingress}
Although there are a few studies of ingress filtering, most studies of spoofing focus on egress filtering. What can be inferred from egress filtering on igress filtering and vice versa?

In their recent measurement of ingress and egress filtering \cite{luckie2019network} conclude that filtering of inbound spoofed packets is less deployed than filtering of outbound packets, despite the fact that spoofed inbound packets pose a threat to the receiving network. \cite{korczynski2020don} analysed the networks from Spoofer and open resolver projects and found that 74\% of the networks that do not filter outbound spoofed packets, do not filter inbound spoofed packets. A more recent study \cite{korczynski2020closed} of 515 ASes found that ingress filtering of inbound spoofed packets is more widely deployed than egress filtering of outbound packets.

%Out of 515 ASes, 81 are vulnerable to outbound spoofing while as many as 114 and 207 ASes are fully and partially vulnerable to inbound spoofing. It says "the results suggest that when network operators are familiar with the concept of SAV, they tend to secure traffic leaving their networks."
The correlation between egress and ingress filtering in previous work shows that the measurements of ingress filtering also provide a lower bound on the number of networks that enforce egress filtering of spoofed outbound packets. Therefore our results on networks that do not enforce ingress filtering imply that at least as many networks do not perform egress filtering.

\subsection{Measurements of Spoofability}
Measurements of networks that filter spoofed packets in the Internet was previously done using {\em network traces} or using {\em vantage points}. We summarise the results of the previous studies in Table \ref{tab:related:work}, and briefly explain them below. %Vantage points were volunteers as well as networks with faulty or misconfigured servers. Studies of network traces were done over traceroute logs and passively by analysing traffic at IXP. We summarise the results of the previous studies in Table \ref{tab:related:work}, and briefly explain them below. %Some of the studies below measure ingress filtering, the others egress filtering. We compare to both types of studies since they all are important for ensuring security and stability of the Internet infrastructure and there are large efforts to mitigate both problems.

\begin{table*}[ht!]
\centering
 \begin{tabular}{|| c | c | c | c | c | c | c | c | c ||} 
 \hline
 Study & Coverage & Spoofable &  Type & Year & Longitudinal & Reproducible & Scalable \\ [0.5ex] 
  & (scanned ASes) & ASes  &   &  &  &  & \\%[0.5ex] 
 \hline\hline
 Spoofer Project \cite{beverly2005spoofer} & 202 of 18,000 (1.1\%) & 52  & Egress & 2005 & \cmark &\xmark & \xmark\\ 
   Spoofer Project \cite{beverly2013initial} & 1,586 of 44,000 (3.6\%) & 390 & Egress & 2013 & \cmark &\xmark & \xmark\\ 
   Misconfigured servers \cite{kuhrer2014exit} & 2,692 of 48,000 (5.6\%) & 870 & Egress & 2014 & \xmark &\cmark & \cmark\\
     Traceroute \cite{lone2017using} & 1,780 of 56,000 (3.2\%) & 703 & Ingress & 2017 & \xmark &\xmark & (\cmark)\\
     IXP traces \cite{lichtblau2017detection} & 700 of 56,000 (1.3\%) & 393 & In \& Eg & 2017 & \xmark &\xmark & \xmark \\
  Amazon Turk Spoofer Project \cite{lone2018using} & 784 of 56,000 (1.4\%) & 48 & Egress & 6w. in 2017 & \cmark & \xmark & \xmark\\  
  Spoofer Project \cite{luckie2019network} & 5,178 of 66,000 (7.8\%) & 1,631 & In \& Eg & 2019 & \cmark & \xmark & \xmark \\
  \textbf{SMap} & 63,522 of 70,468 (90\%) & 51,046 & Ingress & 2019-21 & \cmark & \cmark & \cmark\\ [1ex]
%{\bf SMap & 30563 of 65689 (46.5\%) & 13780 & - & 2019-20 & \cmark & \cmark}\\[1ex] 
 \hline
 \end{tabular}
 \caption{Comparison between SMap and other studies.}
 \label{tab:related:work}
%\vspace{-20pt}
\end{table*}

%%% VANTAGE POINTS: VOLUNTEERS %%%
{\bf Vantage Points.} Measurement of networks which do not perform egress filtering of packets with spoofed IP addresses was first presented by the Spoofer Project in 2005 \cite{beverly2005spoofer}. The idea behind the Spoofer Project is to craft packets with spoofed IP addresses and check receipt thereof on the vantage points operated by the {\em volunteers}, i.e., participants who run a ``spoofer'' software provided by the authors. %The code uses raw sockets to generate UDP packets with spoofed source IP addresses sent to the hosts operated by the volunteers. The spoofability is established by measuring the fraction of arriving packets from the clients to the measurement servers. 
Based on the data collected by the Spoofer Project many reports were published providing statistics on the deployment of egress filtering in the Internet \cite{beverly2009understanding,beverly2013initial,lone2018using,luckie2019network}; we list the statistics in Table \ref{tab:related:work}. 
%In 2005 they identified that clients in 73 ASes could spoof packets. 
%In 2014 \cite{zhang2014mismanagement} performed a study to check for correlation between mismanagement and maliciousness of networks, among others also analysed egress filtering with the Spoofer Project dataset of 2987 ASes, and found 638 of them not to enforce egress filtering. 

The downside of this approach is that the Spoofer Project requires users to download, compile and execute a software - which also needs administrative privileges to run - once per measurement. This requires not only technically knowledgeable volunteers that agree to run untrusted code, but also networks which agree to operate such vantage points on their premises. %The failures of the Spoofer Project are mostly related to failures to open raw sockets and send spoofed packets on some clients. 
%Although an optimal setup for spoofing measurements, the vantage points provide a limited coverage of the Internet.
\cite{huz2015experience} argues that extending the limited coverage of the Spoofer Project is difficult: the operators are unlikely to volunteer or conduct measurements that could leak a negative security posture of their networks, including lack of support of BCP38 \cite{rfc2827}. Hence, \cite{huz2015experience} propose that the most viable method to measure filtering of spoofed packets in more networks is by crowd-sourcing.  %Their own experiment failed since Amazon at that time did not allow running experiments that required workers to download software. A few years later Amazon relaxed their policies, eventually allowing experiments which require to download software. 
In 2018 \cite{lone2018using} performed a one-time study of the Spoofer Project by renting a 2,000 EUR crowd-sourcing platforms with workers that executed the Spoofer software over a 6 weeks period. Their measurements included additional 342 ASes which were not covered by the Spoofer Project previously. Crowd-sourcing studies, in addition to being expensive, are also limited by the networks in which workers are present and do not provide longitudinal and repetitive studies that can be validated and reproduced. 

In a recent longitudinal data analysis by the Spoofer Project \cite{luckie2019network} the authors observed that despite increase in the coverage of ASes that do not perform ingress filtering in the Internet, the test coverage across networks and geo-locations is still non-uniform.

%%% VANTAGE POINTS: MISCONFIGURED %%%
Closely related to {\em volunteers} is the vantage points measurements with {\em faulty or misconfigured servers}. \cite{mauch2013open} noticed that some DNS resolvers do not change the source IP addresses of the DNS requests that they forward to upstream resolvers and return the DNS responses using the IP addresses of the upstream resolvers - a problem which the authors trace to broken networking implementations. \cite{kuhrer2014exit} used this observation to measure egress filtering in networks that operate such misconfigured DNS resolvers. %The idea is that the DNS resolver receives a request, performs the resolution and sends a response but without changing the IP address of the response packet to its own address. Since the response is sent from IP address which does not belong to the AS on which the DNS server is located the authors conclude that the network does not perform egress filtering. 
Such measurements are limited only to networks which operate DNS servers with broken networking implementations: out of 225,888 networks that \cite{kuhrer2014exit} measured, they could find such DNS servers only in 870 networks. 

Since the Open Resolver and the Spoofer Projects are the only two infrastructures providing vantage points for measuring spoofing - their importance is immense as they facilitated many research works analysing the spoofability of networks based on the datasets collected by these infrastructures. Nevertheless, the studies using these infrastructure, e.g., \cite{huz2015experience,luckie2019network}, point out the problems with the representativeness of the collected data of the larger Internet. Both projects (the Spoofer and the Open Resolver) acknowledged the need to increase the coverage of the measurements, as well as the challenges for obtaining better coverage and stable vantage points.

% either ingress or egress.
%%% NETWORK TRACES: ACTIVE vs. PASSIVE %%%
{\bf Network Traces.} To overcome the dependency on vantage points for running the tests, researchers explored alternatives for inferring filtering of spoofed packets. A recent work used loops in traceroute to infer ability to send packets from spoofed IP addresses, \cite{lone2017using}. 
%The idea is that if the traceroute leaves the stub AS with the IP address of the vantage point initiating the traceroute and reaches the upstream provider then the provider does not perform ingress filtering. If the upstream provider's border router performs filtering, it should block the traceroute packet when it arrives from the stub AS, as the packet has a source address not belonging to the stub AS. 
This method detects lack of ingress filtering only on provider ASes (i.e., spoofable customer ASes cannot be detected). The study in \cite{lone2017using} identified loops in 1,780 ASes, which is 3.2\% of all the ASes, and 703 of the ASes were found spoofable. Although a valuable complementary technique for active probes with vantage points, this approach has significant limitations: in the absence of loops ingress filtering cannot be inferred, alternately a forwarding loop in traceroute does not imply absence of filtering at the edge, since a loop resulting from a transient misconfiguration or routing update can occur anywhere in the network. Therefore, to identify a lack of ingress filtering reliably one needs to detect a border router and, more importantly, the traceroute loops need to be reproduced - a difficult problem in practice. Furthermore, reproducing or validating the dataset after some time is virtually impossible as the odds for failures rapidly increase. Running traceroutes is also challenging: black-holes in traceroutes, whereby the routers do not respond to probes or when routers have a limit for ICMP responses, are common in Internet \cite{marchetta2016and}.

\cite{lichtblau2017detection} developed a methodology to passively detect spoofed packets in traces recorded at a European IXP connecting 700 networks. The limitation of this approach is that it requires cooperation of the IXP to perform the analysis over the traffic and applies only to networks connected to the IXP. Allowing to identify spoofing that defacto took place, the approach proposed in \cite{lichtblau2017detection} misses out on the networks which do not enforce filtering but which did not receive packets from spoofed IP addresses (at least during the time frame in which the traces were collected).

%does not allow to run a {\em targeted} evaluation whether some network in question filters spoofed traffic. Specifically, this approach allows to identify networks that received packets with spoofed IP addresses during the time period during which the analysis was made. However, it misses out on the networks which do not enforce filtering but which did not receive packets from spoofed IP addresses (at least during the time frame in which the traces were collected).

A range of studies analysed network traces for ingress filtering using IP address characteristics \cite{moore2006inferring,barford2006toward,chen2008spatial,czyz2014taming,dainotti2013estimating}, or by inspecting on-path network equipment reaction to unwanted traffic, \cite{yao2014passive}. In addition to a limited coverage, the studies do not support longitudinal and repeating data collection and analysis, and cannot be reproduced as they do not make the datasets of their studies public.

%% file: tools.tex
\section{Scanning for Spoofable Networks}\label{sc:tools}

%SMap measures ingress filtering against remote services, hence it needs to identify the services in the networks that it received for a scan, and then to perform the scan.

\subsection{Dataset} SMap architecture consists of two parts: dataset scan and ingress filtering scan. The dataset scan collects the popular services using two methods: domain-based scan and IPv4 based scan. In IPv4 scan to locate the services SMap probes every IP, checking for open ports that correspond to the services that we need; for instance, port 25 for Email, 53 for DNS, 80/443 for Web. To reduce the traffic volume of the scan, instead of probing each IP address for target ports, SMap enables also query of the input domains for services. For every domain, it queries the IP and hostname of the services, e.g., (A, MX) for Email server, A for Web server, (A, NS) for name server. %The algorithms, describing the steps needed for collecting the servers according to domains scan and IPv4 scan are illustrated in Appendix, in Algorithms \ref{algo:dataset-scan-domain} and \ref{algo:dataset-scan-ipv4} respectively. 
\subsection{Methodology}\label{sc:ipid}\label{sc:method}
% This section needs to be rewritten
The measurement methodology underlying SMap uses active probes, some sent from spoofed as well as from real source IP addresses to popular services on the tested networks. The spoofed source IP addresses belong to the tested networks (similarly to the Spoofer Project \cite{beverly2005spoofer}). The idea behind our methodology is that if the packets with spoofed addresses reach the services in the tested networks, they trigger a certain action. This action can be measured remotely. If the action was not triggered, we conclude that spoofed packets did not reach the service. 

We develop three techniques to detect if networks filter spoofed traffic based on our methodology: DNS lookup, IPID and PMTUD based. Using popular services ensures that our measurements apply to as many Internet networks as possible. 

SMap consists of the orchestrator which coordinates and synchronises the prober hosts. The prober hosts receive the dataset of networks to be scanned for spoofability from the orchestrator. The probers then run IPID, PMTUD and DNS lookup tests against the services on the dataset list. SMap applies one test at a time for each AS in the dataset. Each successful test indicates that packets from a spoofed IP address reached the destination on the target network, implying that the target AS does not filter spoofed packets. On the other hand, a failed test may indicate that one of the ASes on the path between the probers and the service on the target AS may be filtering spoofed packets.

The results from the tests are stored in the backend database. The GUI displays the results of the measurements at \url{https://smap.cad.sit.fraunhofer.de}. We next explain each measurement technique. In our measurements in Section \ref{sc:data} we compare the success and applicability of each technique.

\subsection{IPID} 
Each IP packet contains an IP Identifier (IPID) field, which allows the recipient to identify fragments of the same original IP packet. The IPID field is 16 bits in IPv4, and for each packet the Operating System (OS) at the sender assigns a new IPID value. There are different IPID assignment algorithms which can be categorised as: random and predictable. Predictable category uses either a global counter or multiple counters per designation IP address, such that the counter is incremented in predictable quotas. Random category selects each IPID value at random from a pool of values. 

Recent work showed that even TCP traffic gets fragmented under certain conditions \cite{DBLP:conf/anrw/DaiSW21}. Fragmentation has long history of attacks which affect both the UDP and TCP traffic \cite{kent1987fragmentation,cns:frag:dns,shulman2014fragmentation}.

{\bf Methodology.} We use services that assign globally incremental IPID values. The idea is that globally incremental IPID [RFC6864] \cite{rfc6864} values leak traffic volume arriving at the service and can be measured by any Internet host. %\cite{danezis2008covert} used a globally incremental IPID for exchanging covert communication between two hosts, by encoding bits in traffic volume.
%designing a covert channel between two hosts Alice and Bob. The idea is the following: the parties identify a server that assigns globally incremental IPID values. Alice encodes information by sending a set of packets to that server (e.g., $n$ packets correspond to encoding of bit 1, and $2\times n$ correspond to encoding of bit 0) while Bob probes the IPID values on the server to infer the bit that was encoded by Alice. 
%We follow a similar idea to design a technique for inferring ingress filtering. 
Given a server with a globally incremental IPID on the tested network, we sample the IPID value (send a packet to the server and receive a response) from the IP addresses controlled by us. We then generate a set of packets to the server from spoofed IP addresses, belonging to the tested network. We probe the IPID value again, by sending packets from our real IP address. If the spoofed packets reached the server, they incremented the IPID counter on the server - an event which we infer when probing the value from our real IP address the second time. 

The challenge here is to accurately probe the increments rate of the IPID value (caused by the packets from other sources not controlled by us), in order to be able to extrapolate the value that will have been assigned to our second probe from a real source IP. This allows us to infer if the spoofed packets incremented the IPID counter.

{\bf Identifying servers with global IPID counters.} We send packets from two hosts (with different IP addresses) to a server on a tested network. We implemented probing over TCP SYN, ping and using requests/responses to Name servers and we apply the suitable test depending on the server that we identify on the tested network. If the responses contain globally incremental IPID values - we use the service for ingress filtering measurement with IPID technique. We located globally incremental IPID in $63.27\%$ of the measured networks. There are certainly more hosts on networks that support globally incremental IPID values, yet our goal was to validate our measurement techniques while keeping the measurement traffic low - hence we avoided scanning the networks for additional hosts and only checked for Web, Email or Name servers with globally incremental IPID counters via queries to the tested domain.
\begin{figure}
\centering
\includegraphics[width=0.45\textwidth]{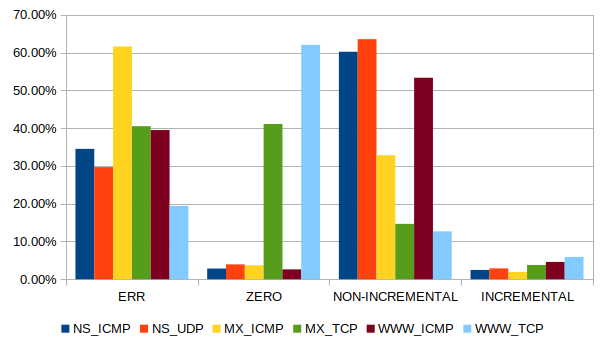}
%\vspace{-10pt}
\caption{IPIDs on servers in dataset.}
\label{fig:measure-ipid-compare}
%\vspace{-15pt}
\end{figure}

Statistics of IPID values distribution among tested servers are plotted in Figure~\ref{fig:measure-ipid-compare}. When ICMP is filtered, it results in ERROR, when run with TCP, the IPID values are often zero (i.e., ZERO IPID in graph) in Figure \ref{fig:measure-ipid-compare}. To improve coverage of the IPID technique we merge the ICMP\&TCP and ICMP\&UDP results for each server in our measurements.

\begin{figure} [t!]
  \begin{center}
   \includegraphics[width=\columnwidth]{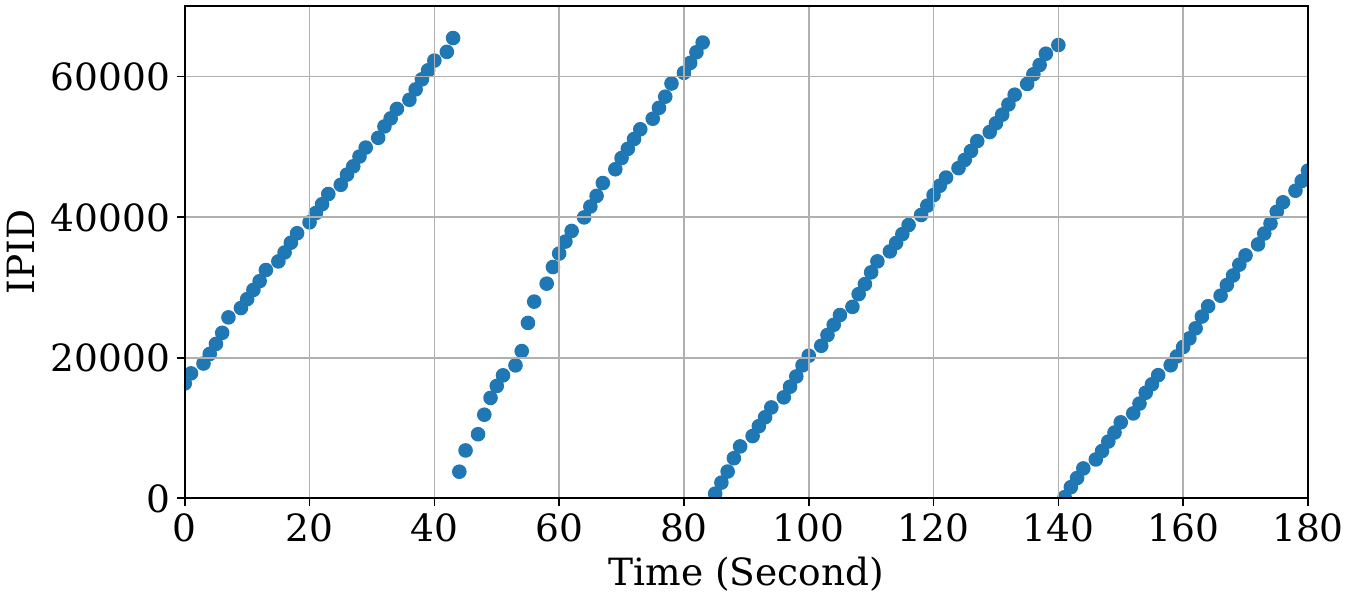}
   %\vspace{-15pt}
		\caption{IPID of Name server 69.13.54.XXX during 180sec.}
		\label{fig:incremental_ipid}
		%\vspace{-15pt}
  \end{center}
\end{figure}

{\bf Measuring IPID increment rate.} The traffic to the servers is stable and hence can be predicted, \cite{wessels2003wow}. We validate this by sampling the IPID value at the servers which we use for running the test. One example evaluation of IPID sampling on one of the busiest servers is plotted in Figure \ref{fig:incremental_ipid}. In this evaluation we issued queries to a Name server at {\tt 69.13.54.XXX} during three minutes, and plot the IPID values received in responses in Figure \ref{fig:incremental_ipid} - the identical patterns demonstrate predictable increment rates. Which means that the traffic to the server arrives at a stable rate.

%We calculate the range of the candidate IPID values that will be t time $t$... TBD

{\bf Accuracy of IPID measurements.} The IPID techniques are known to be difficult to leverage, requiring significant statistical analyses to ensure correctness. Recently, \cite{ensafi2014detecting,pearce2017augur} developed statistical methods for measuring IPID. However, in contrast to our work, the goal in \cite{ensafi2014detecting,pearce2017augur} is different - they use IPID to measure censorship and have additional sources of inaccuracy, which do not apply to our measurements: (1) the measurements are applied against client hosts, which results in significantly higher noise than our measurements against servers - the clients move between networks, change IP addresses, the clients are located behind intermediate devices, such as Network Address translators (NAT) and firewalls - which also prevents direct measurements; (2) inaccuracies in geolocation tools, which do not apply to our study since we do not need to know the location to measure ingress filtering, (2) additional network mechanisms (anycast, rerouting, traffic shaping, transient network failures). All these can only cause us to classify the server as not 'testable', but do not impact 'spoofable' outcomes. Furthermore, the IPID measurement methods in prior workss use TCP-RST packets to increment IPID, which are often blocked in firewalls. In contrast, we use packets which are not blocked such as DNS queries or TCP-SYN.

{\bf Inferring spoofing.} We use the following components: the prober at IP address {\tt 7.7.7.7} and a server at IP address {\tt 1.2.3.7} that uses globally incremental IPID, illustrated in Figure \ref{fig:ipid}. 
Using the prober at {\tt 7.7.7.7}, we measure the value of the IPID and the rate at which IPID increments. We use linear regression with Ordinary Least Square (OLS) method to estimate the relation between IPID and timestamp $t$. %According to Figure~\ref{fig:incremental_ipid}, 
Since IPID is incremental, it holds: $IPID = a * t + b + \epsilon, \epsilon \sim N(0, \sigma^{2})$

We send $N$ probes to {\tt 7.7.7.7} (in step (1)). % in Figure \ref{fig:ipid-new}). 
With $N$ probes, we can estimate $a$, $b$ and $\sigma$ using OLS method in step (2). In step (3) in Figure \ref{fig:ipid} we send a set of $M = 6 * \sigma$ packets from a spoofed source IP address {\tt 1.2.3.6} (belonging to the probed network). In step (4) at time \( T_{M+N+1} \) we sample the IPID value \( Z = IPID^{real}_{M+N+1} \) from the server from the prober's real IP address {\tt 7.7.7.7} - this is needed in order to receive the response. We check the IPID value $Z$ in step (5) in Figure \ref{fig:ipid}. 
Taking the linear regression model into consideration, we can calculate \( IPID^{esti}_{M+N+1} \) at time \( T_{M+N+1} \). If the $M$ spoofed packets are filtered, according to \textit{3-sigma rule}, there is a $99.73\%$ possibility that: $IPID^{esti}_{M+N+1} - 3 * \sigma \leq Z \leq IPID^{esti}_{M+N+1} + 3 * \sigma$.
However, if the spoofed packets are not blocked, a.k.a. there is no ingress filtering, the IPID counter should have an additional increment of $M$. Thus \( Z > IPID^{esti}_{M+N+1} + 3 * \sigma \), which is also \( Z > IPID^{esti}_{M+N+1} + M / 2 \). %\todel{P-code for IPID technique is shown in Appendix, Figure \ref{fig:method-ipid-process}}.

%\subsubsection{Detailed Process}
\begin{figure}[ht!]
  \centering
  %\vspace{-10pt}
  \includegraphics[width=0.3\textwidth]{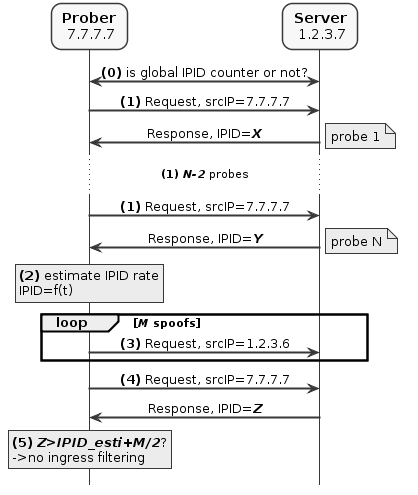}
  %\vspace{-10pt}
  \caption{Sequence diagram for IPID technique.}
 % \label{fig:method-ipid-process}
  \label{fig:ipid}
\end{figure}

\ignore{
\begin{figure}[ht!]
\centering
%\vspace{-10pt}
\includegraphics[width=0.3\textwidth]{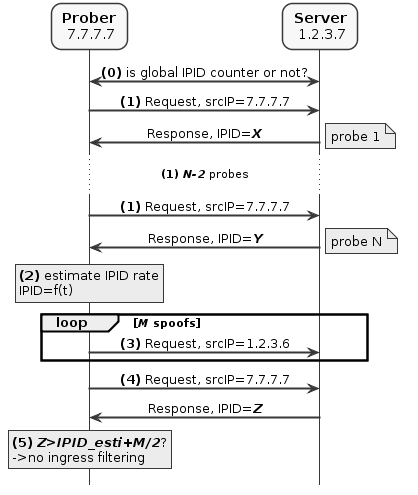}
%\vspace{-10pt}
\caption{Sequence diagram for test with IPID technique.}
%\vspace{-10pt}
\label{fig:method-ipid-process}
\label{fig:ipid}
\end{figure}
}
%of IPID test is showed in Fig.~\ref{fig:method-ipid-process}. As a pre-requirement of IPID test, we must make sure if target Server is using globally sequential IPID or not. We need at least two VMs with different IPs to test that. In our case, we use a Master VM and a Servant VM. 

We define outcomes of a test with IPID technique as {\em spoofable, applicable, non-applicable, N/A}; see Table \ref{tab:method-ipid-case}. The IPID technique is not applicable if the IPID counter is constant zero or if the IPID counter is not globally incremental.
\begin{table}[hbt!]
\centering
\resizebox{0.45\textwidth}{!}{%
\begin{tabular}{ |c|c|c|c| }
 
 \hline
  {\bf Category} & {\bf IPID} & {\bf PMTUD} & {\bf DNS} \\
 %                &            &             & {\bf Lookup}\\
 \hline
  Spoofable
   & \begin{tabular}{@{}c@{}}no filtering \\ \end{tabular}
   & \begin{tabular}{@{}c@{}}no filtering \\ \end{tabular}
   & \begin{tabular}{@{}c@{}}no filtering \\ \end{tabular}
  \\
 \hline
  Applicable
   & \begin{tabular}{@{}c@{}}server w/globally \\ incremental IPID\end{tabular}
   & \begin{tabular}{@{}c@{}}host supports\\ PMTUD\end{tabular}
   & \begin{tabular}{@{}c@{}}has DNS\\ server\end{tabular}
  \\
 \hline
  Non-applicable
   & \begin{tabular}{@{}c@{}}random IPID \\ or per-dest IPID\\ or IPID=0\end{tabular}
   & \begin{tabular}{@{}c@{}}(DF$\equiv$0 \& MF$\equiv$0) or\\ (DF$\equiv$1 or MF$\equiv$1) \&\\ no change\end{tabular}
   & \begin{tabular}{@{}c@{}}\end{tabular}
  \\
  \cline{1-3}
  N/A
  & \begin{tabular}{@{}c@{}}host unreachable \\ or firewall\\or packet loss\\ or load balancer\end{tabular}
  & \begin{tabular}{@{}c@{}}host unreachable or \\ misconfigured service \\ or firewall\\or packet loss\end{tabular}
  & \begin{tabular}{@{}c@{}}no DNS \\ server \\ found\end{tabular}
  \\
  \hline
\end{tabular}}
\caption{Outcomes of tests.}% with IPID, PMTUD and DNS techniques.}
\label{tab:method-ipid-case}
%\vspace{-20pt}
\end{table}

\ignore{
\begin{figure}[ht!]
\centering
%\vspace{-5pt}
\includegraphics[width=0.3\textwidth]{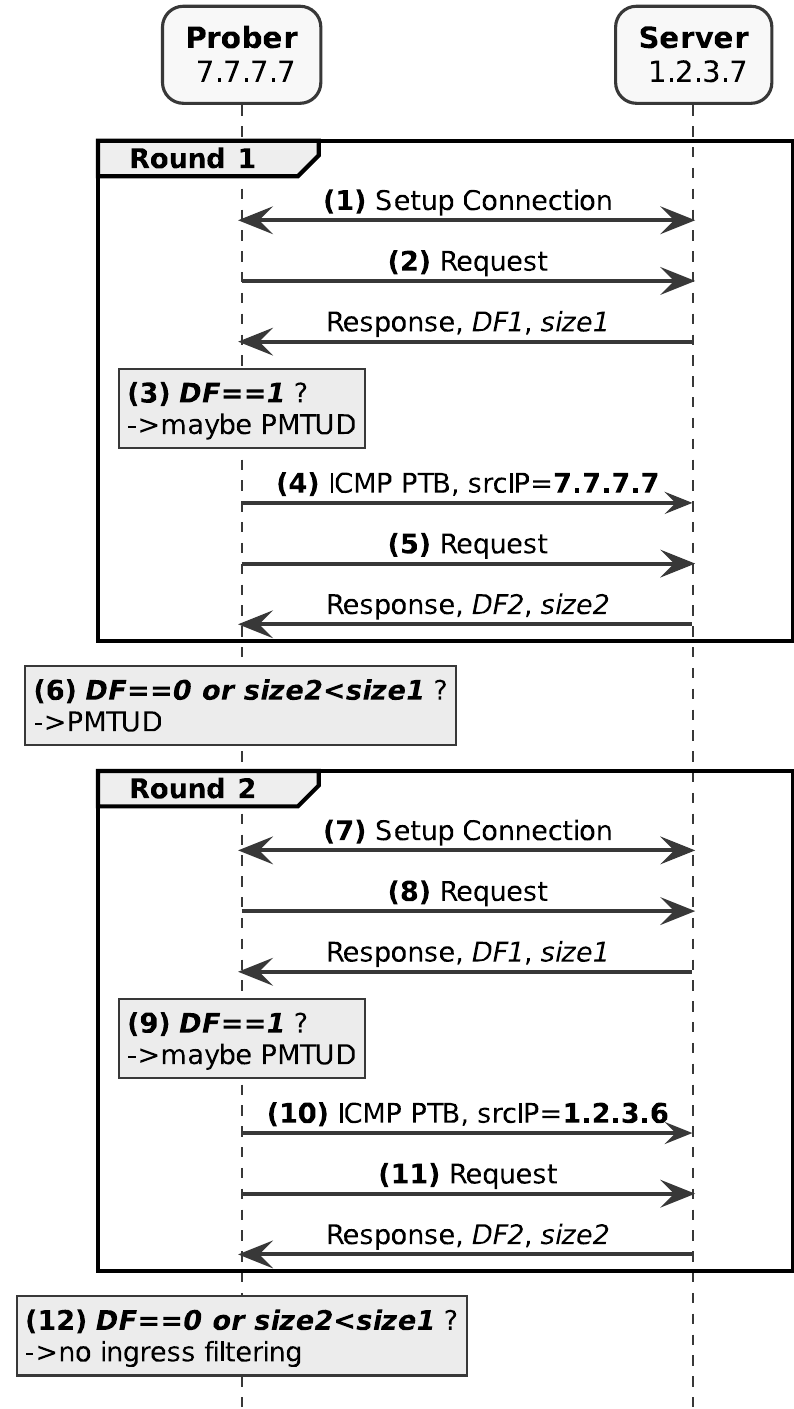}
%\vspace{-10pt}
\caption{Sequence diagram for test with PMTUD technique.}
%\vspace{-5pt}
\label{fig:method-pmtud-process}
\end{figure}
}

\subsection{PMTUD} Path Maximum Transmission Unit Discovery (PMTUD) determines the MTU size on the network path between two IP hosts. The process starts by setting the {\tt Don't Fragment (DF)} bit in IP headers. Any router along the path whose MTU is smaller than the packet will drop the packet, and send back an ICMP Fragmentation Needed / Packet Too Big (PTB). The payload of the ICMP packet contains the IP header and the first 8 bytes of the original packet that triggered the error as well as the MTU of the router that sent the ICMP message. %; see ICMP packet format in Appendix, Figure \ref{fig:pkt_icmp_ptb}. 
After receiving an ICMP PTB message, the source host should either reduce its path MTU appropriately or unset the DF bit.

A study of CAIDA datasets in 2017 found 3M ICMP fragmentation needed packets sent by routers in the Internet, with about 1K routers sending ICMP error message with next hop MTU of less than 500 Bytes \cite{gohring2018path}.
\begin{figure}[htbp!]
\centering
%\vspace{-5pt}
\includegraphics[width=0.3\textwidth]{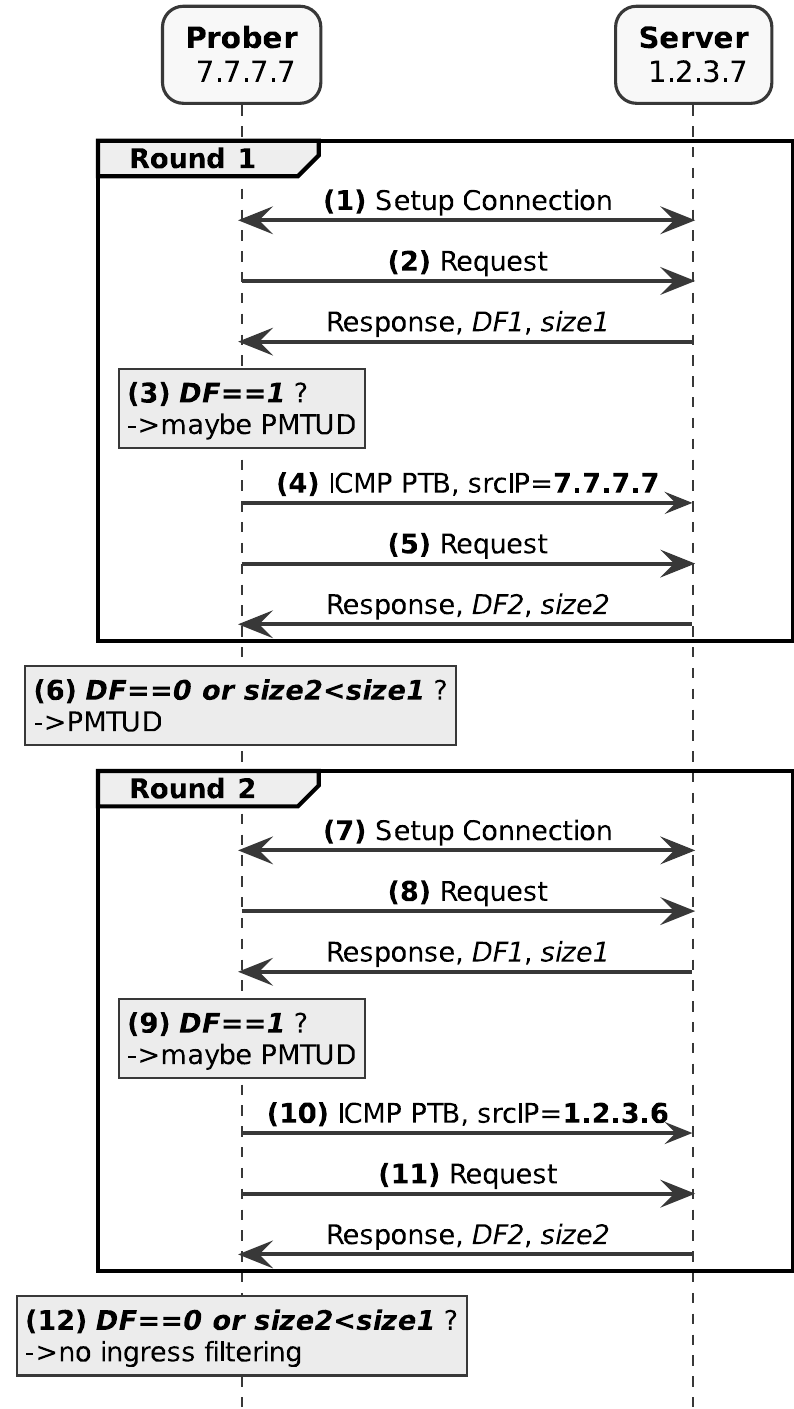}
%\vspace{-10pt}
\caption{Sequence diagram for PMTUD technique.}
%\vspace{-5pt}
\label{fig:method-pmtud-process}
\end{figure}
\ignore{
\begin{figure} [ht]
  \begin{center}
   \includegraphics[width=0.43\textwidth]{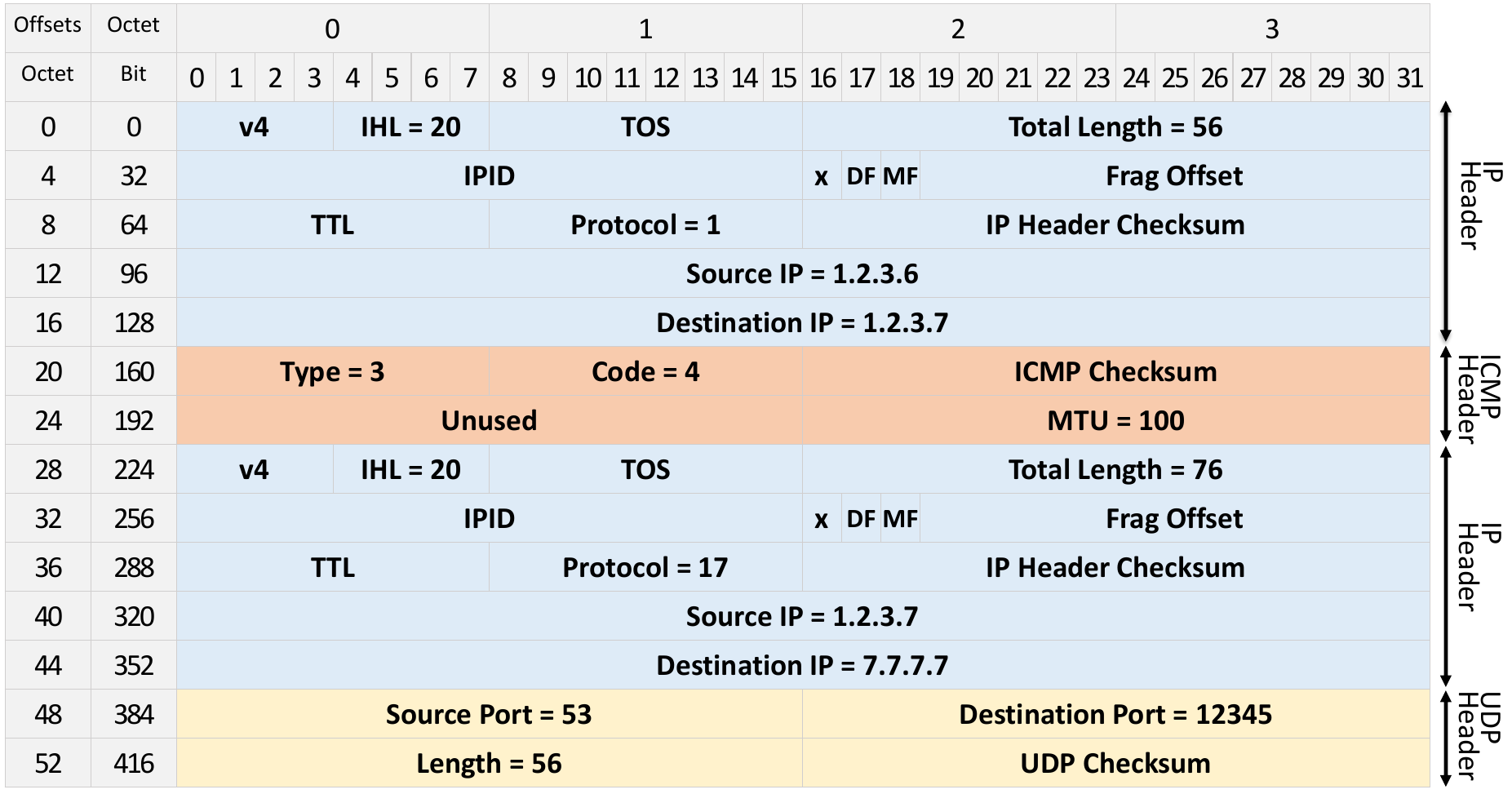}
		\caption{ICMP fragmentation needed packet from prober at 7.7.7.7 to server at 1.2.3.7 indicating an MTU of 100 bytes with spoofed source IP}
		\label{fig:pkt_icmp_ptb}
		%\vspace{-10pt}
  \end{center}
\end{figure}
}
%\subsubsection{Methodology.} 

{\bf Methodology.} The core idea of the Path MTU Discovery (PMTUD) based tool is to send the ICMP Packet too Big (PTB) message from a spoofed source IP address, belonging to the tested network, and in the 8 bytes payload of the ICMP to insert the real IP address belonging to the prober. If the network does not enforce ingress filtering, the server will receive the PMTUD message and will reduce the MTU to the IP address specified in the first 8 bytes of the ICMP payload. We first probe the MTU to a service on the tested network, then send ICMP PTB from a spoofed IP address. If the packet arrives at the service, it will reduce the MTU to our prober, and we will identify this event in the next packet from the service - this event implies that the tested network does not apply ingress filtering.

%\subsubsection{Identifying servers that support PMTUD} 
{\bf Identifying servers that support PMTUD.} We measured networks that support PMTUD (i.e., do not filter ICMP Fragmentation Needed (Type 3, Code 4) messages), and found that $85.92\%$ of the tested networks support PMTUD.

%\subsubsection{Inferring spoofing.} 

{\bf Inferring spoofing.} The PMTUD test is illustrated in Figure~\ref{fig:method-pmtud-process}. We establish a TCP connection to a server on the tested network. Then we send Request1 and receive Response1. If DF bit is not set, the server does not support PMTUD. Otherwise, we send an ICMP PTB with smaller MTU. Following that, we request again and get Response2. If \( DF_1 == 1 \ and \ ( DF_2 == 0\  or\  size_2 \leq size_1 ) \), the server supports PMTUD. Now we can proceed to test if ingress filtering is enforced. We spoof an ICMP PTB with smallest MTU, using server's neighbour IP as source IP address. Once that is done, we make another request. The server is not protected by ingress filtering if following condition applies:
$size_3 \leq size_2 \  or \  (DF_2 == 1 \ and \ DF_3 == 0)$.

%and spoof an ICMP PTB indicating our host has a smaller MTU. If the network does not enforce ingress filtering, the server reacts to our PTB message according to PMTUD value in the payload of the ICMP packet.

We define outcomes of a test with PMTUD technique as {\em spoofable, applicable, non-applicable, N/A}; see rightmost column in Table \ref{tab:method-ipid-case}. 
\ignore{
\begin{table}[hbt!]
\centering
\begin{tabular}{ |c|c| }
 \hline
  Category & Case \\
 \hline
  Spoofable
  & \begin{tabular}{@{}c@{}}host accepts \\ spoofed packets\end{tabular}
  \\
 \hline
  Testable
  & \begin{tabular}{@{}c@{}}host supports PMTUD\end{tabular}
  \\
 \hline
  Non-Testable
  & \begin{tabular}{@{}c@{}}DF==0 and MF==0\end{tabular}
  \\
 \hline
  Non-Testable
  & \begin{tabular}{@{}c@{}}DF==1 and no change \\ or MF==1 and no change\end{tabular}
  \\
  \hline
  N/A
  & \begin{tabular}{@{}c@{}}host unreachable \\ or service misbehave \\ or filtered by firewall\end{tabular}
  \\
  \hline
  N/A
  & \begin{tabular}{@{}c@{}}packet loss\end{tabular}
  \\
  \hline
\end{tabular}
\caption{Cases of PMTUD Side Channel}
\label{tab:method-pmtud-case}
\end{table}

\begin{figure}
\centering
%\vspace{-5pt}
\includegraphics[width=0.4\textwidth]{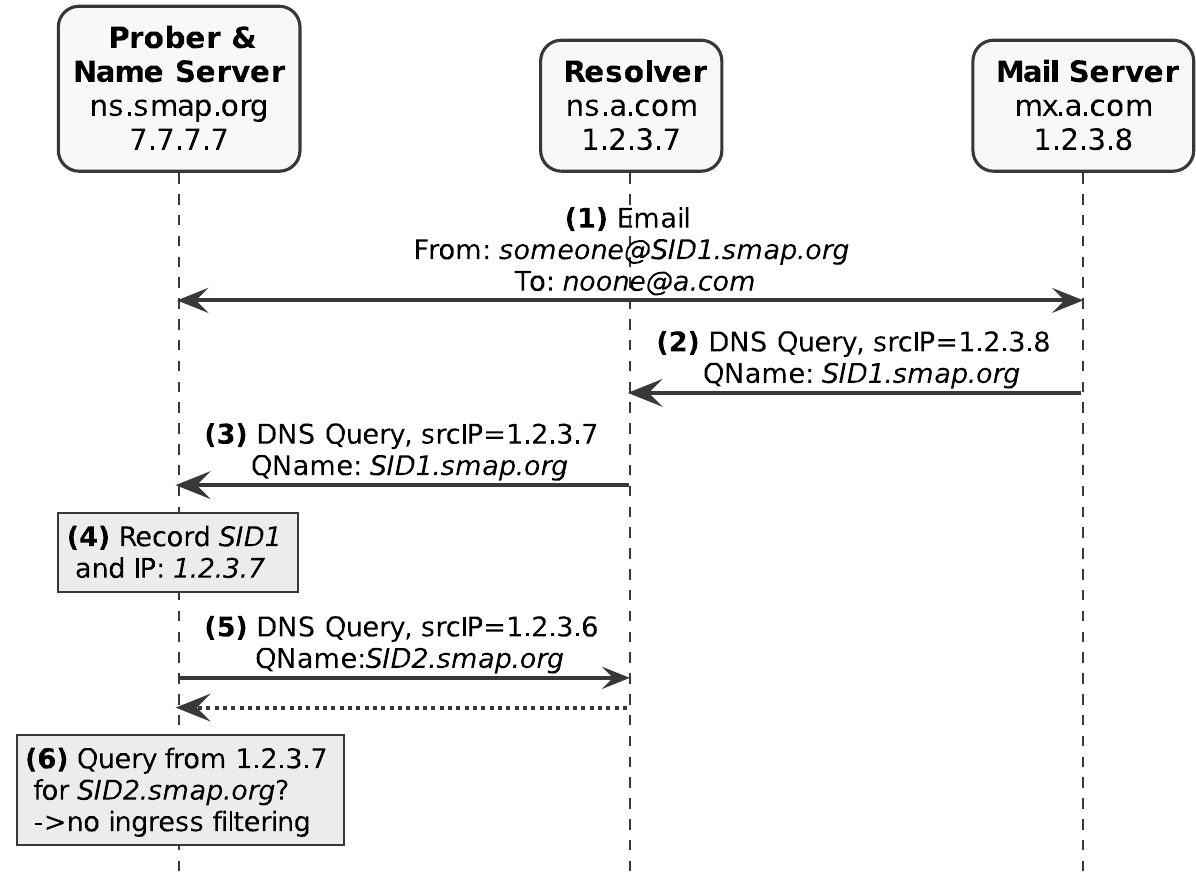}
%\vspace{-10pt}
\caption{Sequence diagram for test with DNS lookup technique.}
%\vspace{-10pt}
\label{fig:method-dns-process}
\end{figure}
}
\subsection{DNS Lookup}
DNS provides lookup services to networks. Upon receiving a DNS request, the resolver performs the lookup of the requested domain name and returns the response with the requested record.

%\subsubsection{Methodology.} 

{\bf Methodology.} We send a DNS request to the tested network from a spoofed IP address belonging to the tested network. If the network does not enforce ingress filtering, the request will arrive at the DNS resolver on that network. A query from a spoofed source IP address will cause the response to be sent to the IP address from which the request was sent, i.e., the spoofed IP address. Since we do not control the spoofed IP address, we will not be able to observe this event and hence will not be able to infer if the DNS resolver received our request or if the request was filtered due to spoofing. To obtain insights into the traffic arriving at the resolver in the tested network we utilise the payload of the DNS request: the query contains the domain which we own, set up on Name servers that we control. Namely, eventhough the response from the DNS resolver will be returned to the spoofed IP address and will not be received by us, the DNS request will be issued to our Name servers, which is an indication that the DNS resolver on the tested network received our DNS request, sent from spoofed IP address.

%\subsubsection{Identifying DNS resolvers.} 
{\bf Identifying DNS resolvers.} The main challenge here is to locate the DNS resolvers within a domain/network and to trigger a DNS request to our Name servers. We use Email service in the target networks (retrieved via the MX type request in the target domain) to find the DNS resolvers. We send an email to target domain's Email server from one of our unique subdomains with a non-existing recipient set in the destination. This causes the Email server on the tested network to generate a Delivery Status Notification (DSN) error message [RFC3464] to our Email server. To be able to send us the DSN, the Email server will request the resolver on the tested network, to provide it the MX and A/AAAA records of our Email exchanger. At the same time, it may also trigger anti-spam checking, which requests (SPF/TXT, PTR, DKIM, DMARC)-type records in domains under our control. By monitoring the DNS queries at our Name servers, we collect the IP addresses of the resolvers. Using this methodology we identified 49,252 DNS resolvers in 7,141 networks. However, in our regular IPv4 scan, to reduce Email traffic in the Internet, we use the list of servers with UDP port 53 open from Project Sonar as input.

%\subsubsection{Inferring spoofing.}
\begin{figure}[htbp!]
\centering
%\vspace{-5pt}
\includegraphics[width=0.4\textwidth]{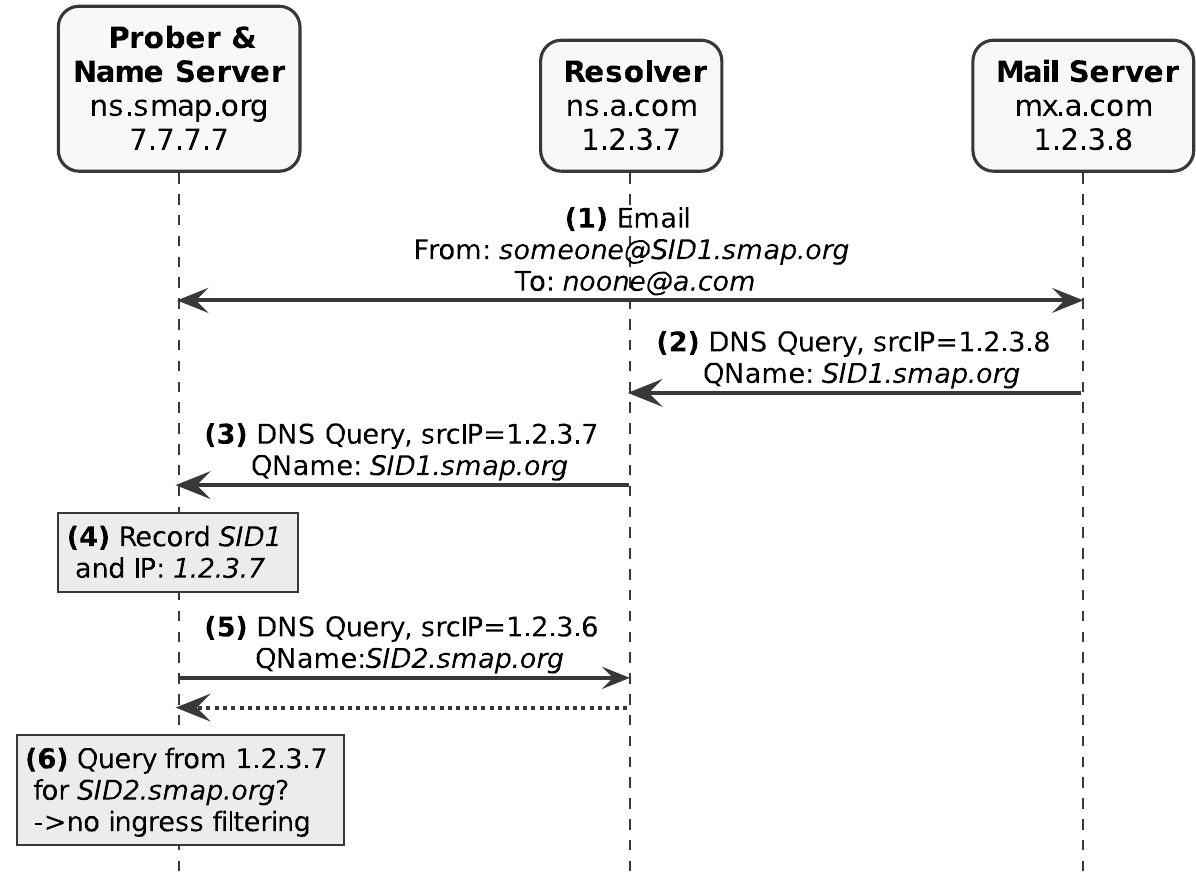}
%\vspace{-10pt}
\caption{Sequence diagram for DNS lookup technique.}
%\vspace{-10pt}
\label{fig:method-dns-process}
\end{figure}
{\bf Inferring spoofing.} Given a DNS resolver at IP {\tt 1.2.3.7}, we send a DNS query to {\tt 1.2.3.7} port 53 asking for a record in domain under our control. The query is sent from a spoofed source IP address belonging to the tested network. We monitor for DNS requests arriving at our Name server. If a query for the requested record arrives from {\tt 1.2.3.7}, we mark the network as not enforcing ingress filtering. The process is illustrated in Figure \ref{fig:method-dns-process}, steps (1-4) locate the IP address of the DNS resolver, and steps (5,6) test for ingress filtering on that network.

%Non-testable means:
%1. none of the name/mail/web servers we find has globally incremental IPID.
%2. none of the name/mail/web servers we find supports PMTUD.
%3. we don't see any resolvers from those ASes.

%\subsection{Volume vs. Accuracy Tradeoff}

%% file: measurements.tex
\section{Internet Measurements}\label{sc:data}
%We fetch Alexa Top-1M list every week and merge it with previous lists. In our latest scan, we have more than 3M domains. For each domain in our list, we get IPs of its name servers via DNS. To acheive that, we first query its NS Resource Records and then query A RRs for each NS in response. Afterwards, we can run IPID test and PMTUD test with that IPs and domain as input. Test s via mail servers or web servers have similar process, just changing the DNS queries.
%Our tools enable the first black-box Internet wide measurements of spoofing ability in the Internet. The first step towards evaluation is to identify the services, in the target networks, against which we will run our tools. We describe our methodology for collecting the services, ensuring sufficient coverage of popular networks while keep the ``scan'' traffic low to prevent loading the Internet and the target networks. We then report on our evaluation results of the tools over the services in popular networks.
% We download a fresh top-1M Alexa list % that we provide in an input to SMaps. We then provide the measurements results of each of the three techniques integrated into SMap over the dataset of networks.

%In our measurement, listed in Table~\ref{tab:measure-stats-domain}, we tested over 3M domains. The domains in which we could not locate any services via DNS queries or if no servers could be reached (e.g., due to a timeout) are marked as ``N/A''. 

\begin{table*}[ht!]
\centering
\begin{tabular}{ |p{2.5cm}|p{1.2cm}|p{1.2cm}|p{1.2cm}|p{1.2cm}|p{2.2cm}|p{1.5cm}|p{1.5cm}|  }
 \hline
  Technique\_Service & \multicolumn{2}{|c|}{Spoofable} & \multicolumn{2}{|c|}{Applicable} & Non-Applicable & N/A & Total ASes\\
 \hline
 IPID\_NS   & 8,752  & 23.07\% & 12,056 & 31.78\% & 25,881 & 25,585 & 63,522 \\
 \hline
 IPID\_MX   & 4,355  & 21.48\% & 6,861 & 33.84\% & 13,416 & 43,245 & 63,522 \\
 \hline
 IPID\_WWW  & 30,963  & 51.83\% & 39,370 & 63.27\% & 22,891 & 2,608  & 63,522 \\
 \hline
 IPID\_ANY  & 32,248  & 56.25\% & 41199 & 67.52\% & 22,853 & 1,299   & 63,522 \\
 \hline
 PMTUD\_NS  & 9,054   & 24.16\% & 11,592 & 30.93\% & 25,885 & 26,045 & 63,522 \\
 \hline
 PMTUD\_MX  & 23,078  & 68.69\% & 27,127 & 80.74\% & 6,471  & 29,924 & 63,522 \\
 \hline
 PMTUD\_WWW & 41,959  & 76.91\% & 47,524 & 87.11\% & 7,034  & 8,964 & 63,522 \\
 \hline
 PMTUD\_ANY & 43,473  & 75.98\% & 49,161 & 85.92\% & 8,053 & 6,308  & 63,522 \\
 \hline
 DNS lookup & 25,407  & 40.00\% & 44,577 & 70.18\% & -     & -     & 63,522 \\
 \hline
 ANY        & 51,046  & 80.90\% & 58,432 & 92.61\% & 4,662  & 428   & 63,522 \\
 \hline
\end{tabular}
\caption{Collected data and analysis per AS view.}
%\vspace{-15pt}
\label{tab:measure-stats-network}
\end{table*}

In this section we report on our Internet-wide measurement of ingress filtering with SMap. Our dataset collection with SMap has been initiated on July 2019 continually collected data over a period of one year, of over 6M domains and an entire IPv4 address block. 

%SMap receives in an input parameter indicating if it should start with domain-scan or with IPv4 address scan. It can also skip the scan phase by receiving a dataset of an already performed scan, as we also did in the measurements in this work.
\subsection{Dataset} SMap first collects the dataset of services.% Our dataset contains services that can be reached through popular Alexa domains. For each of the 6M domains we collect Name servers, Email servers and Web servers. The 6M domains have services hosted in 33,854 ASes with 388,185 Name servers in 23,323 ASes, 687,897 Email servers in 24,735 ASes, and 1,683,496 Web servers in 24,228 ASes. We also collect the services
%{\bf Domain-scan dataset.} We use domains of the popular services according to Alexa. For each of the 6M domains we collect Name servers, Email servers and Web servers. The 6M domains have services hosted in 33,854 ASes with 388,185 Name servers in 23,323 ASes, 687,897 Email servers in 24,735 ASes, and 1,683,496 Web servers in 24,228 ASes. The pseudocode of domain-scan is in Appendix, Algorithm \ref{algo:dataset-scan-domain}.
%{\bf IPv4-scan dataset.} The pseudocode of IPv4-scan that we implemented in SMap is in Appendix, Algorithm \ref{algo:dataset-scan-ipv4}. To avoid scanning the Internet for open ports in the services that we need, 
Our dataset is constructed as follows: we periodically download the entire IPv4 scan from Sonar Project \cite{sonar}. We use the scan results on UDP port 53 as input for Name servers and DNS resolvers, scan data on TCP port 25 for Mail servers and scan results on TCP port 80 for Web servers. Besides, we also make use of forward DNS responses and reverse DNS responses from Sonar Project to help find hostnames of servers. In the latest dataset from Sonar, we have services hosted in 63,522 ASes (Table \ref{tab:measure-stats-network}) with 4,256,598 DNS servers in 38,838 ASes, 16,478,938 Email servers in 38,937 ASes, and 62,455,254 Web servers in 61,535 ASes; see Table \ref{tab:measure-exec-time}.

\ignore{
\begin{algorithm}[ht!]
\caption{IPv4-based Dataset Scan}
\label{algo:dataset-scan-ipv4}
\SetAlgoLined
\KwIn{List: IP}
\KwIn{Opt-Out List: Domain or IP or CIDR}
\KwOut{List: (AS, Domain, IP, Service)}
initialise AS-list\;
\For {each IP} {
  \If {IP in Opt-Out} {
    stop and go to next \textit{IP};
  }
  check open ports for services\;
  \If {no DNS/Mail/Web server} {
    stop and go to next \textit{IP};
  }
  query hostname in reverse DNS\;
  extract domain from hostname\;
  \If {domain in Opt-Out} {
    stop and go to next \textit{IP}\;
  }  
  get AS info from GeoLite2-ASN database\;
  \For {each service} {
    update (AS, Domain, Service, IP) in AS-list\;
  }
}
\end{algorithm} 
}

\begin{table}[t!]
\centering
\begin{tabular}{ |p{1.5cm}|p{1.5cm}|p{1.5cm}|p{1.5cm}| }
 \hline
             & Name Server & Email Server & Web Server \\
 \hline
  \#IPs      & 4,256,598 & 16,478,938 & 62,455,254 \\
 \hline
  \#Blocks   & 697,851  & 748,406   & 3,207,393  \\
 \hline
  \#Prefixes & 229,981  & 217,334   & 542,983   \\
 \hline
  \#ASes     & 38,838   & 38,937    & 61,535    \\
 \hline
%  Preparation & 00:33:40 & 00:24:18 & 00:25:50 \\
% \hline
%  IPID & 00:45:49 & 02:02:31 & 04:52:10 \\
% \hline
%  PMTUD & 00:28:05 & 00:58:40 & 02:41:11 \\
%\hline
\end{tabular}
\caption{Servers in tested networks.}
%\vspace{-15pt}
\label{tab:measure-exec-time}
\end{table}

\ignore{

\subsubsection{Measurement Setup}
 %The execution time of last measurement is displayed in Table~\ref{tab:measure-exec-time}. %Considering our low VPS configuration, our implementation is quite efficient. If we want, we can finish a whole scan in one day. However, for load balance as well as better network utilisation, we prefer to spread the scan over a whole week. 
}

\ignore{
\subsection{IPID}
\begin{figure}
\includegraphics[width=0.5\textwidth]{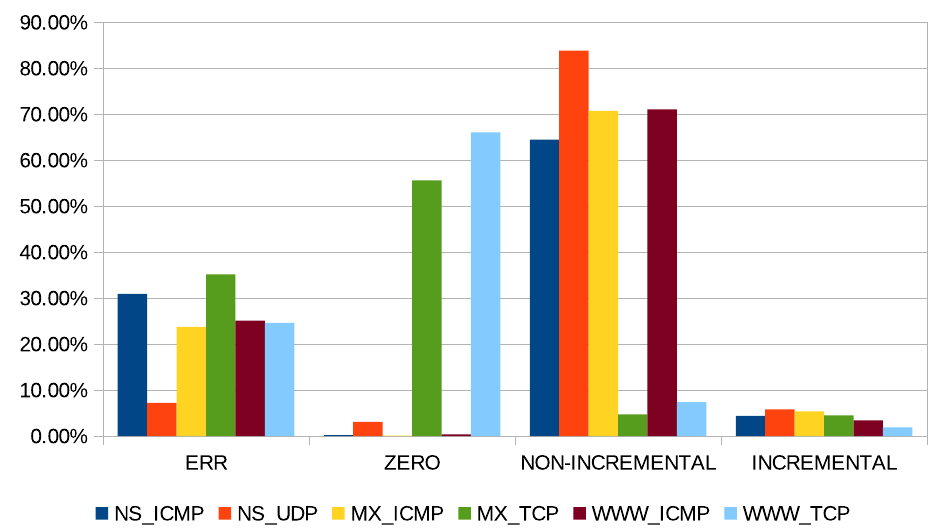}
\caption{Servers IPID measurements.}
\label{fig:measure-ipid-compare}
\end{figure}
We run IPID test against web server, email server and DNS server of each domain on the list. For each server, we use ping both over ICMP and TCP as well as DNS over UDP. Results are showed in Figure~\ref{fig:measure-ipid-compare}. We can see, generally, email servers are more vulnerable than the other two types of servers.

Around 5\% of all the email servers tested are vulnerable to spoofing attack. The number is 1.5\% for web server and 3.5\% for DNS server. One may argue that there are much more web servers than email servers or DNS servers. However, email server even have the highest absolute number of vulnerable servers. We see over 18K vulnerable email servers, while only 13K vulnerable web servers.

 To better cover both cases, we will merge ICMP\&TCP or ICMP\&UDP results for each server in following discussion.
}
%fig:measure-ases-timeseries

\subsection{Ingress Filtering Results} 
\begin{figure}[ht!]
\includegraphics[width=\columnwidth]{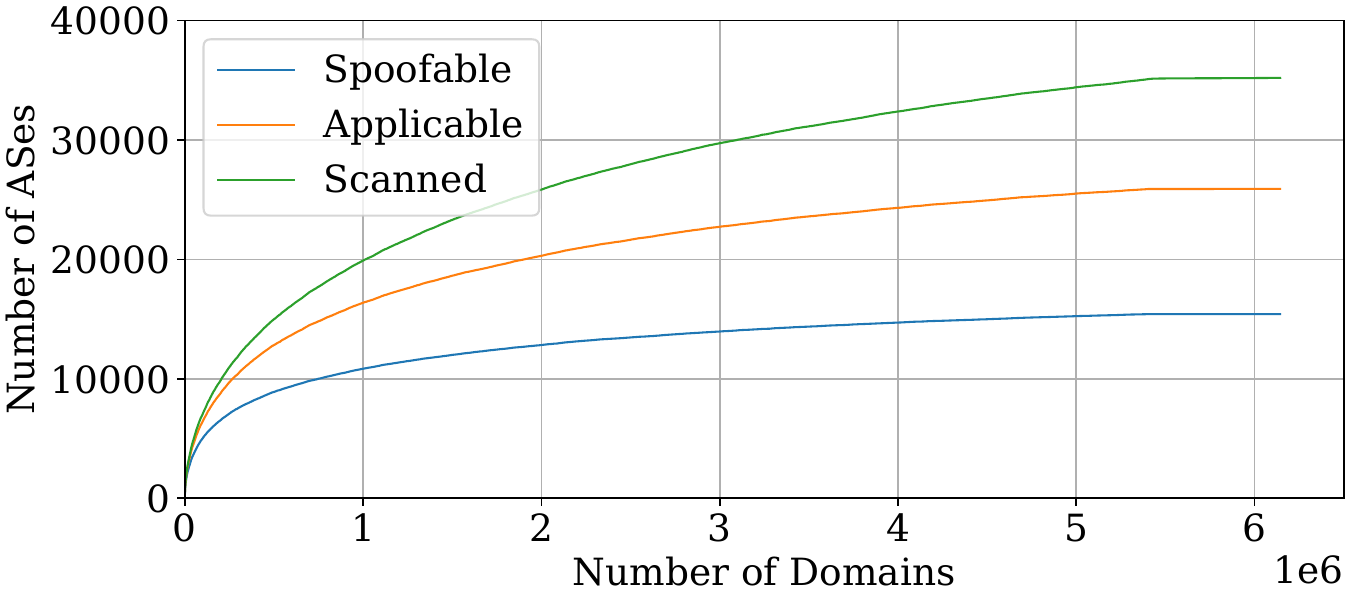}
%\vspace{-15pt}
\caption{As we scan more domains, we cover more ASes and discover more spoofable ASes.}
%\vspace{-10pt}
\label{fig:num_of_ases_in_domains}
\end{figure}

Domain-scan and IPv4-scan both show that the number of spoofable ASes grows with the overall number of the ASes in the Internet, see Figure \ref{fig:measure-ases-timeseries}. Furthermore, there is a correlation between fraction of scanned domains and ASes. Essentially the more domains are scanned, the more ASes are covered, and more spoofable ASes are discovered; see Figure \ref{fig:num_of_ases_in_domains}. This result is of independent interest as it implies that one can avoid scanning the IPv4 and instead opt for domains-scan, obtaining a good enough approximation. This not only reduces the volume of traffic needed to carry out studies but also makes the study much more efficient.

Further, to avoid single point of failure it is recommended that the Name servers of a domain are hosted in multiple networks. This is also our observation when correlating between domains and ASes. %; see Figure \ref{fig:cdf-num-of-ases-in-domain}; and Figure \ref{fig:cdf-num-of-domains-in-as} in Appendix. 
Essentially we find that when testing one domain for each server we can obtain different results, depending on the AS that the server is hosted on. 
%In Appendix, Figure \ref{fig:cdf-num-of-domains-in-as} we plot the correlation between ASes and domains; the plot shows that domains are typically hosted on more than one AS and ASes host multiple domains. 

The results of the ingress filtering measurements with SMap are summarised in Table \ref{tab:measure-stats-network}. The techniques that we integrated into SMap (IPID, PMTUD, DNS lookup) were found applicable to more than 92\% of the measured ASes. Using SMap we identified 80\% of the ASes that do not enforce ingress filtering. %For comparison, in Appendix, Table \ref{tab:measure-stats-domain} we provide data analysis according to domains.%, where the applicability of our techniques is much higher, i.e., more than 90\%. This is due to the fact that the domains have at least one publicly available service, typically name server.
In what follows we compare the effectiveness of the techniques, explain causes for false negatives and failures.
In the rest of this section we explain and analyse the applicability of our results and the success of the different techniques, discuss errors and compare to the results in previous studies.
%In Appendix, Section \ref{asc:comparison}, we use a one day BGP RIS dump from the RouteView Project \cite{routeviews} to map the spoofable ASes to the prefixes announced in the Internet. Our results in Appendix, Table \ref{tab:measure-stats-prefix} demonstrate that more than 352K of the announced prefixes do not enforce ingress filtering.

%(instead of ASes as in Table \ref{tab:measure-stats-network}). Not surprisingly the applicability of our techniques is much higher, i.e., more than 90\%, since domains have at least one publicly available service, typically name server.

\begin{figure}[t!]
\includegraphics[width=\columnwidth]{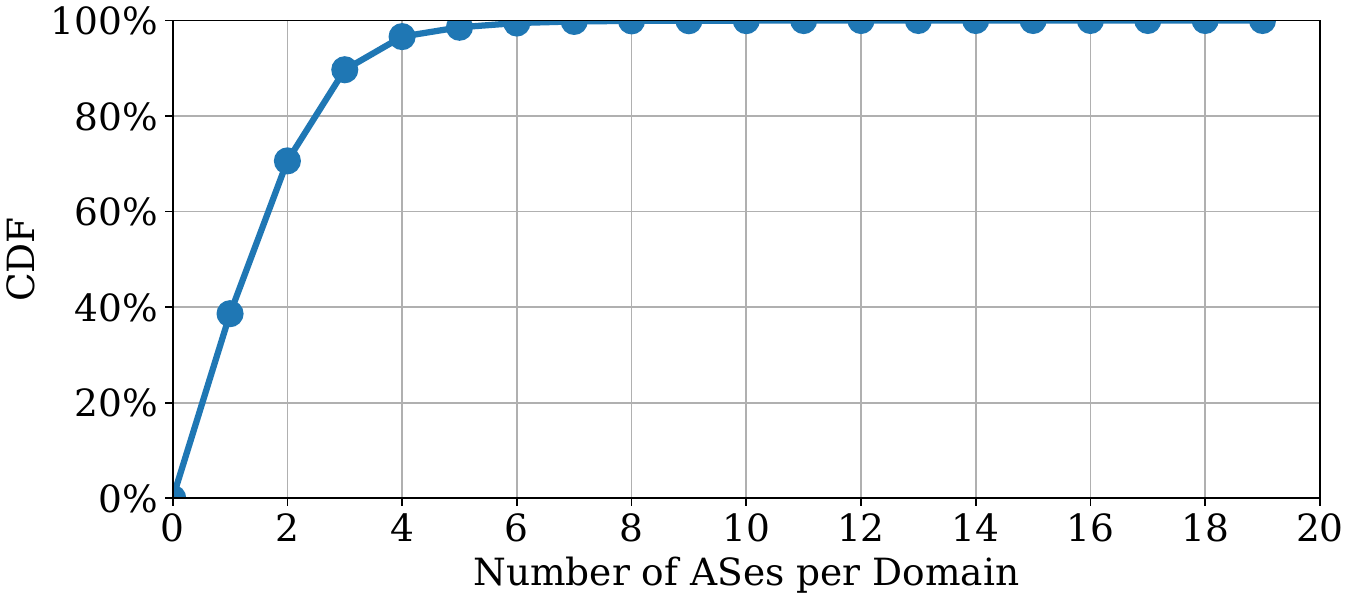}
%\vspace{-15pt}
\caption{Fraction of domains hosted in multiple ASes. We check how many ASes host services of one domain: 70\% of the domains are hosted in one or two ASes.}
%\vspace{-10pt}
\label{fig:cdf-num-of-ases-in-domain}
\end{figure}
\subsection{Applicability and Success} 
As can be seen in Table \ref{tab:measure-stats-network} the most applicable technique is PMTUD against Web servers, which applies to a bit more than 87\% of the ASes, yielded the highest fraction of spoofable ASes. This is not surprising, since %it is expected that firewalls should block port 53 from the Internet to the IP address of the resolver, making the test appear as ``not successful'' even if the firewalls do not actually defacto apply ingress filtering, while 
the number of web servers is much larger than the others and it is recommended not to block ICMP to Web servers to allow for path MTU discovery. %Table \ref{tab:measure-stats-domain}, in Appendix, lists the measurement results according to domains - also here the most applicable test is via DNS lookup while PMTUD technique against Web servers yielded the highest ratio of spoofable domains.

\ignore{ %%% TO CHECK %%%% DNS LOOKUP NOT MOST APPLICABLE
    \begin{figure}[ht!]
\includegraphics[width=0.45\textwidth]{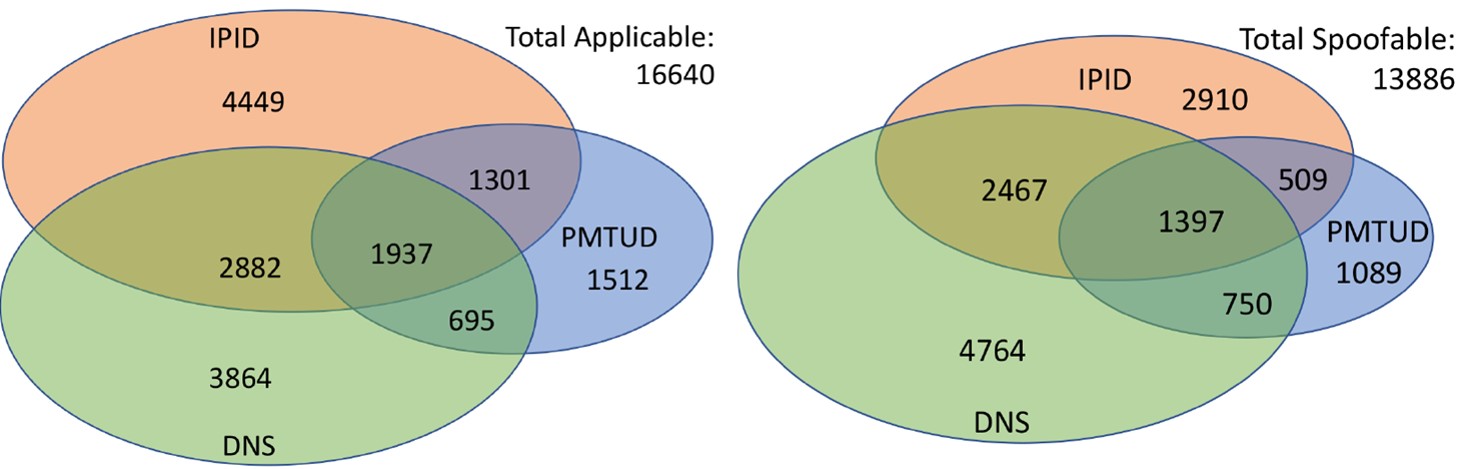}
%\vspace{-5pt}
\caption{Applicability of techniques to ASes (left) and spoofability inference by different techniques (right).}
%\vspace{-10pt}
\label{fig:measure-applicable-by-technique}
\end{figure}

\begin{figure}[ht!]
\includegraphics[width=0.3\textwidth]{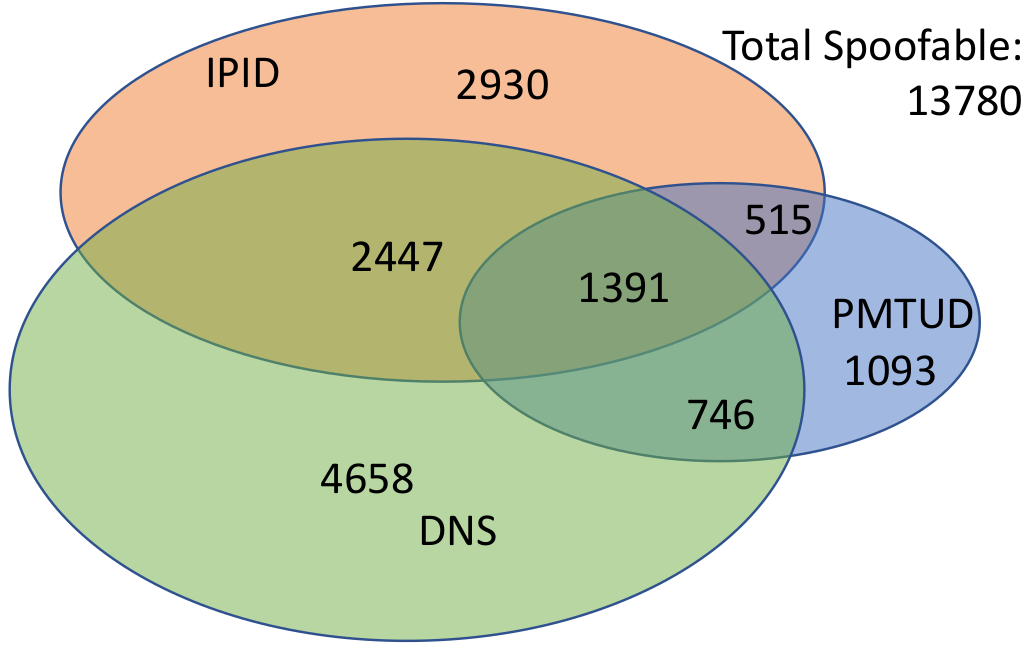}
%\vspace{-15pt}
\caption{Spoofability inference by difference techniques.}
%\vspace{-15pt}
\label{fig:measure-spoofable-by-technique}
\end{figure}
}

    We next compare the success and applicability of tests with PMTUD and IPID techniques against Email, Name and Web servers. In order to compare the effectiveness of the PMTUD and IPID measurement techniques as well as their applicability, we define the spoofable and applicable rates, as follows:
\[Rate_{spoofable} = \frac{N_{spoofable}}{N_{total} - N_{NA}}, Rate_{applicable} = \frac{N_{applicable}}{N_{total} - N_{NA}}\]

\begin{figure}[t!]
\includegraphics[width=0.5\textwidth]{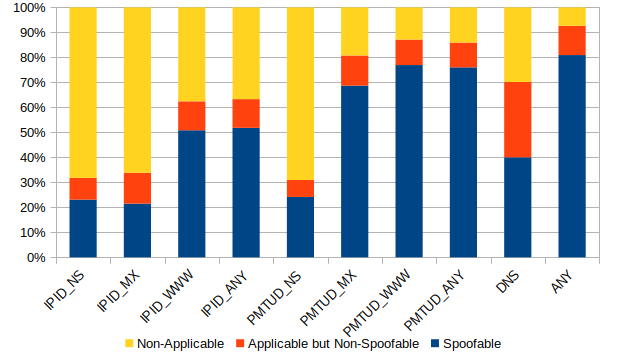}
%\vspace{-15pt}
\caption{Coverage of the measurement techniques.} 
%\vspace{-15pt}
\label{fig:measure-method-compare}
\end{figure}

The spoofable rate reflects the fraction of the networks found not to apply ingress filtering and the applicable rate means applicability of the test technique. The coverage of each of the three techniques for different types of servers (Web, Name, and Email) is plotted in Figure \ref{fig:measure-method-compare}.

\begin{figure}[ht!]
\includegraphics[width=0.47\textwidth]{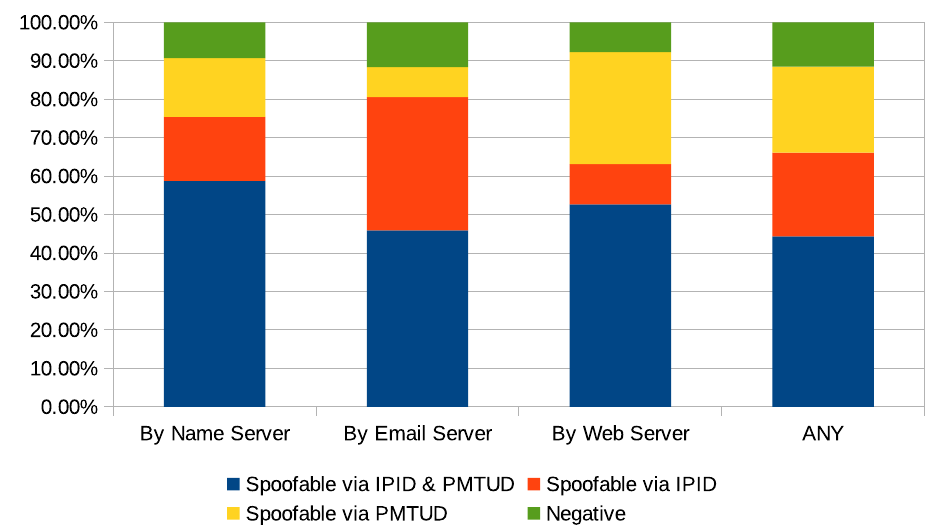}
%\vspace{-10pt}
\caption{Comparison of spoofability via IPID and PMTUD.}
%\vspace{-15pt}
\label{fig:measure-compare-spoofable}
\end{figure}

 Figure \ref{fig:measure-method-compare} shows that PMTUD technique (listed as ``PMTUD\_ANY'' in Figure \ref{fig:measure-method-compare}) has a better test rate than either of the IPID and DNS tests, which indicates that PMTUD is still widely supported. Between the other two, DNS test has a slightly higher applicability than IPID test, which shows that globally sequential IPID is less supported now. In Figure \ref{fig:measure-compare-spoofable} we similarly see that the fraction of spoofable networks that can be fonud through IPID and PMTUD is higher than when measured with the other methodologies; Figure \ref{fig:measure-compare-spoofable} plots the networks found spoofable via IPID vs PMTUD excluding "N/A" networks.

\ignore{
Even though PMTUD test applies to more networks than IPID test, IPID test reports more networks as spoofable. %; see plot in Figure \ref{fig:measure-compare-applicable}. 
Especially when we consider the hit rate \( \frac{N_{spoofable}}{N_{testable}} \), PMTUD test has over 83\%, which is much higher than the other two tests. An interesting finding is that over half of the networks which have a Name server with globally sequential IPID counter are vulnerable to IP spoofing.
}

%In general, tests against Name servers have a higher applicability rate than the tests with Email or Web servers, regardless of which technique was used (IPID or PMTUD). The reason is twofold: first, every domain has at least one Name server while it is not guaranteed that each domain has a Web or an Email server. Second, DNS is a fundamental protocol that any user and service on the Internet needs, in order to locate other services. Indeed, almost all the networks where tests against Name servers were applicable, but against Email or Web not, it was due to the fact that the network did not operate Email/Web servers. To test this, we select a random sample of networks, among those in which the tests against Email and Web did not apply but against Name server applied; a scan shows that ports 80/443 and 25 on those networks are not open. Furthermore, we find that when a Name server is not available (``N/A''), both Email and Web servers cannot be tested, either. This also results in much higher N/A outcomes for tests against Email and Web servers as opposed to Name servers. 
In general, tests against Web servers have a higher applicability rate than the tests with Email or DNS servers, regardless of which technique was used (IPID or PMTUD). The number of Web servers is much larger than the others. It is much easier to setup a Web server than Email server or DNS server. Considering that DNS servers and Email servers are more likely to be hosted by providers, they also have higher probability to get new system updates. Furthermore, we find that when a Web server is not available (``N/A''), both Email and DNS servers cannot be tested, either. This also results in much higher N/A outcomes for tests against Email and DNS servers as opposed to Web servers. 
%showing percentage excluding "N/A" networks, we can say among those testable networks (allowing testing without encountering any error), Email servers have highest applicability rate. 

\begin{figure}[htbp!]
    \centering
    \includegraphics[width=0.2\textwidth]{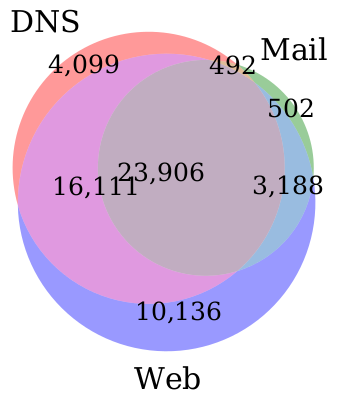}
    \includegraphics[width=0.2\textwidth]{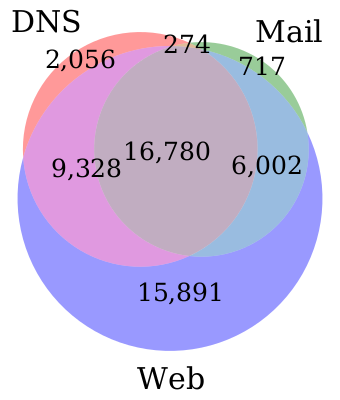}
    \caption{Number of Applicable (left) and Spoofable (right) ASes according to service type.}
    \label{fig:as-by-service}
\end{figure}

The higher applicability of the tests against web servers also correlates with a higher number of spoofable networks. 
%In Figures \ref{fig:measure-applicable-by-service-dns-merged} and \ref{fig:measure-spoofable-by-service-dns-merged}, 
In Figure \ref{fig:as-by-service}, we show the relationships between the applicability of SMap measurement techniques to different services and the overlap between them.

\ignore{
\begin{figure}[htbp!]
    \centering
    \includegraphics[width=0.5\textwidth]{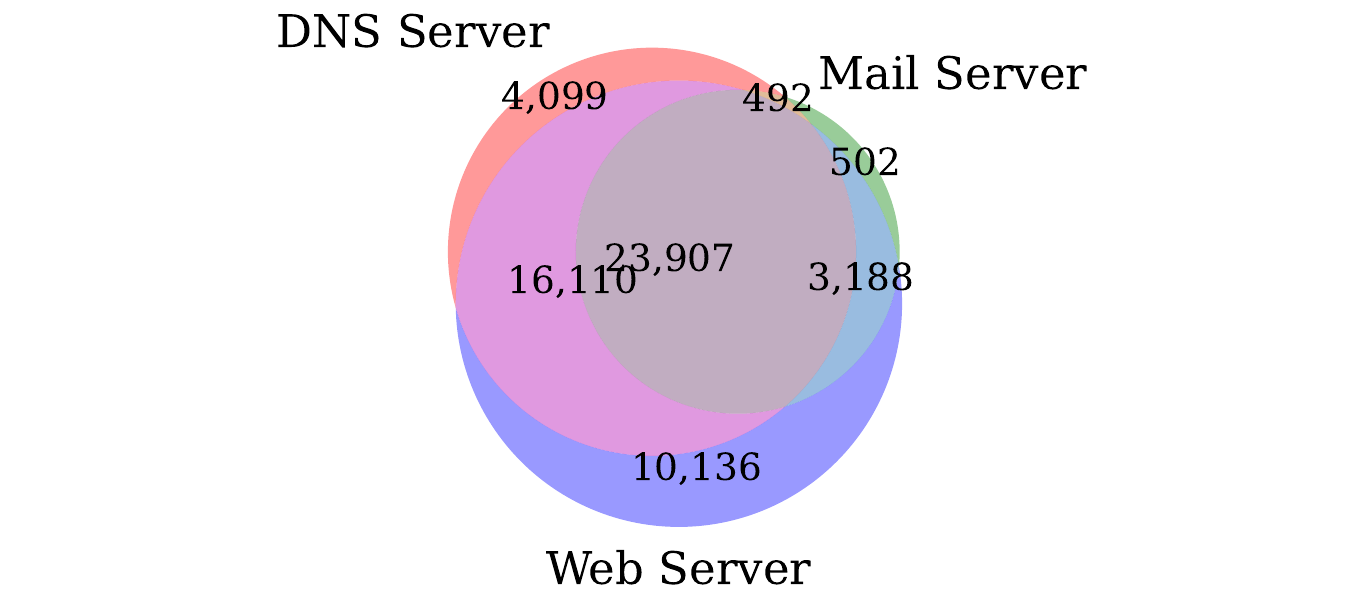}
    \caption{Applicable ASes according to service type.}% (Name Server and Resolver merged).}
    \label{fig:measure-applicable-by-service-dns-merged}
\end{figure}

\begin{figure}[htbp!]
    \centering
    \includegraphics[width=0.5\textwidth]{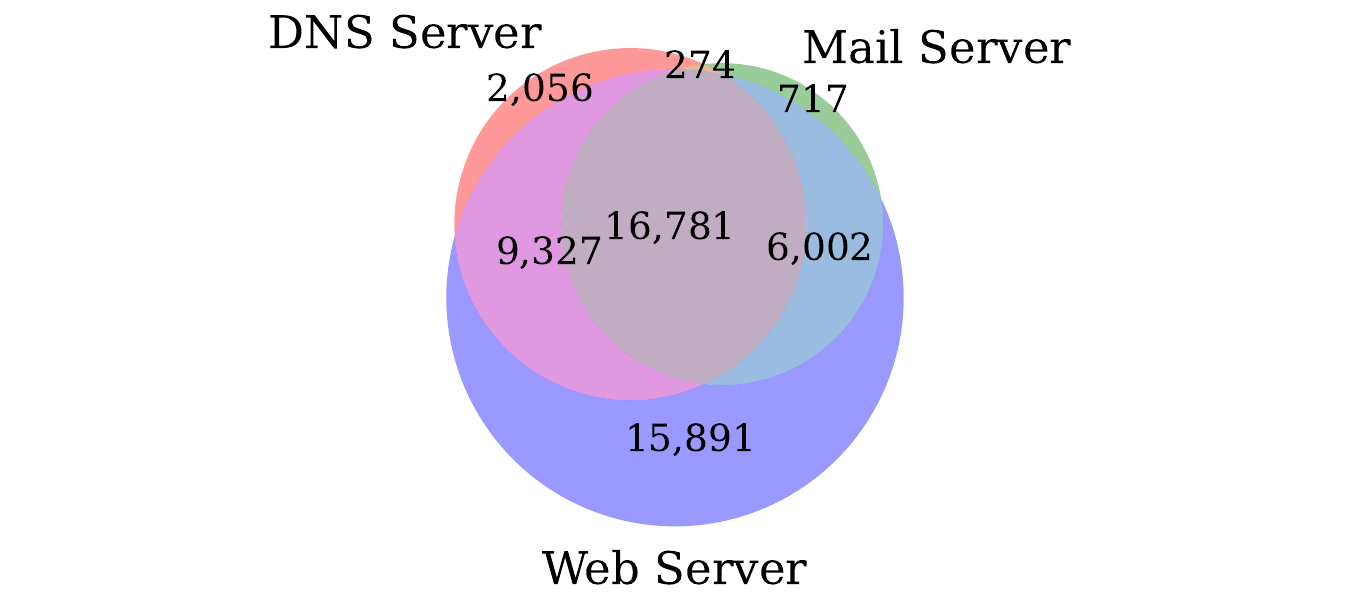}
    \caption{Spoofable ASes according to service type.}% (Name Server and Resolver merged).}
    \label{fig:measure-spoofable-by-service-dns-merged}
\end{figure}
}

%Taking the high test rate of name server test into consideration, no wonder that tests of name servers discover more domains vulnerable to IP spoofing. Remembering that tests of name servers also have a much shorter execution time, as showed in Tab.~\ref{tab:measure-exec-time}, tests against name servers via DNS over UDP have better cost-performance-ratio than that of mail servers and web servers. 
\ignore{
\begin{figure}[ht!]
\includegraphics[width=0.47\textwidth]{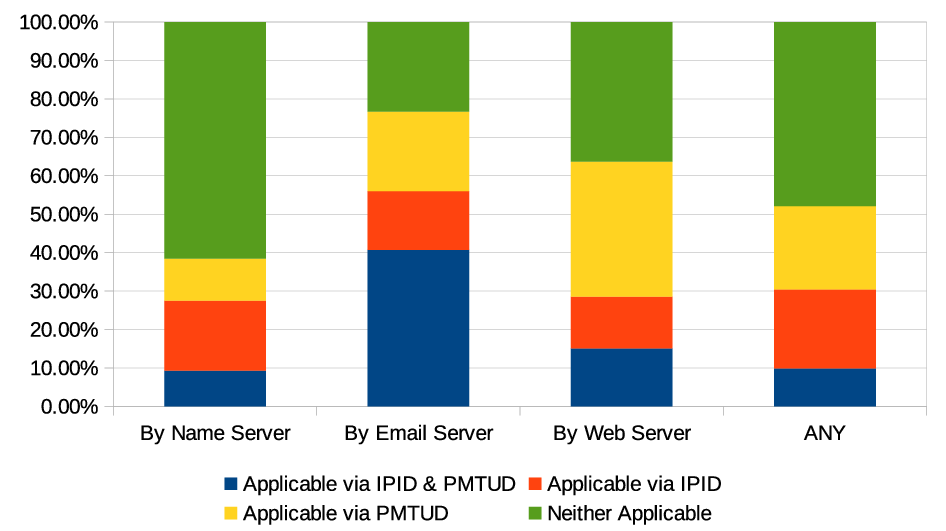}
%\vspace{-5pt}
\caption{Comparison of applicability of IPID and PMTUD.}
%\vspace{-15pt}
\label{fig:measure-compare-applicable}
\end{figure}
 }

%\vspace{-5pt}
\subsection{Errors}
We define the result of SMap evaluation successful (i.e., true positive) if at least one of the three tests outputs that the tested network does {\em not} filter spoofed packets: either the IPID value on the server in the tested network was incremented as expected (IPID test) or we receive a query at our domain (DNS test) or the server on the tested network reduced the MTU of the packets sent to us (PMTUD test). When either of the three techniques provides a positive result, we mark the network as {\em not filtering}. 

SMap does not make mistakes when reporting a network as not filtering. However, it can have false negatives: when the scan does not report network as not filtering when a network does not filter spoofed packets.

\subsubsection{No False Positives}

Our techniques are not susceptible to false positives, that is, classification of the tested network as filtering spoofed packets when in fact it does not do so. This is a side effect of our methodology - only when spoofing is not filtered will the ``test action'' be triggered. 

{\bf IPID technique.} When spoofing is not filtered the counter on the server will be incremented - which is the test action. At the probing phase the counter's value will equal or large than the expected value after the increment phase. The %GPS encoding and
repeated measurements ensure that we do not accidentally interpret noise (i.e., packets from other sources to the same server) as lack of ingress filtering.

{\bf DNS technique.} When spoofing is not filtered the DNS resolver on the tested network will receive a DNS request from a spoofed IP address to our domain. Hence a query at our domain is the test action that spoofed packets are not filtered.

{\bf PMTUD technique.} Reduction of the MTU of the packets sent from the test server to our network is the action which indicates that spoofing filtering is not enforced.

\subsubsection{False Negatives}
False negatives in our measurements mean that a network that does not perform filtering of spoofed packets is not marked as such. We next list the causes of false negatives for each of our three techniques. Essentially the false negatives cannot be resolved, and therefore our measurement results of networks that enforce ingress filtering introduce a a lower bound. The networks that we classify as those that do not apply ingress filtering - definitely allow packets from spoofed IP addresses into the network. The networks which were not classified as ``not enforcing ingress filtering'', could still be ``not enforcing ingress filtering'', but this cannot be determined using our techniques.

{\bf IPID technique.} Load balancing can introduce a challenge in identifying whether a given network enforces ingress filtering. As a result of load balancing our packets will be split between multiple instances of the server, hence resulting in low IPID counter values. There are different approaches for distributing the load to different instances, e.g., random or round robin, which makes it impossible to identify whether a ``load-balanced-server'' is on a network which applies ingress filtering or not.

Anycasted server instances can also introduce a challenge in inferring ingress filtering enforcement. We identified such cases by performing traceroutes to the server.

{\bf DNS technique.} Firewalls, blocking incoming packets on port 53, would as a result generate a similar effect as ingress filtering on our servers: we would not receive any DNS requests to our domain. However, such a setting does not indicate that the tested network actually performs ingress filtering. 

{\bf PMTUD technique.} Firewalls are often configured to block ICMP packets. In such case the evaluation result is similar as when a tested network does not enforce ingress filtering: our PMTUD packets will be blocked by the firewall, but not because they originate from an IP address that belongs to the tested network but because the firewall blocks ICMP packets. This case can be identified by sending ICMP PMTUD packets from an IP address that does not belong to the network. If the ICMP packets are not blocked (but were blocked when the packets were sent from a spoofed IP address) then the network does not block ICMP packets and does enforce IP spoofing filtering. On the other hand if the packets are blocked then one cannot determine if the blocking   is done because of ICMP or because of filtering of spoofed IP addresses.

\ignore{
\subsubsection{Improving Accuracy}
The accuracy of the measurements with SMap can be improved by identifying more servers to which our techniques apply, this however requires generating more traffic. Specifically, in this work our goal is to evaluate the effectiveness of domains-scan by resolving the MX, NS and A records (that correspond to Email, Name and Web servers) and to run the tests against them, providing a lower bound on the effectiveness of our proposal, while keeping the traffic low. Nevertheless, in every AS there are typically multiple hosts which operate (Web, Email or Name) servers on ports 80/443, 25 and 53, in addition to the services that appear in the zonefile of the domain. To identify more servers, a scan of the AS for ports 53, 25 and 80/443 is required. %With multiple servers the probability for servers, to which our techniques apply. %, increases, as well as the probability for locating services in different subnets, which may or may not be enforcing ingress filteringbe differently configured, namely, even if one subnet of the AS applies ingress filtering, it can be that the others do not. %Detecting multiple servers in an AS allows to cover multiple subnets of the AS. %We leave it as a future work to analyse the tradeoff between the traffic needed 
}
%\vspace{-10pt}
\subsection{Comparison with Other Measurements}
To understand the effectiveness of our methodologies we compare the results of our measurements with the active measurements of ingress filtering performed by the CAIDA Spoofer Project. These include two types of measurements: {\em using traceroute} and {\em using agents}. The spoofer project is the only measurement study that makes the datasets from their scans available online. The traceroute approach and the agents approach are the only two other active measurements of enforcement of ingress filtering (see Related Work Section \ref{sc:works}). We crawled all the 217,917 session reports in 2019 of CAIDA Spoofer Project. These included 2,867 ASes with Spoofer Project agents, and 2,500 ASes with Spoofer Project traceroute loops (total of 5,367 ASes). Using our methodologies we measured 63,522 ASes, which is substantially more than the previous studies all together. We compare between our results and the other two methodologies below. %We covered much more ASes than the traceroute and agents approaches together. We compare between our results and the other two methodologies below. 
%We found 9122 ASes to be spoofable among the total number of ASes tested by the Spoofer project using traceroutes and using agents.

% in how many of the networks our techniques do not apply.

%\vspace{-10pt}
{\bf Traceroute Active Measurements.} We analyse the datasets from the traceroute measurements performed by the CAIDA Spoofer Project within the last year 2019, \cite{lone2017using}. The measurements identified 2,500 unique loops, of these 703 were provider ASes, and 1,780 customer ASes. The dataset found 688 ASes that do not enforce ingress filtering. Out of 688 ASes found with traceroutes by the Spoofer Project, we could not test 4 ASes (none of our tests applied) and 36 ASes were not included in our tests (those ASes could not be located from domain names - due to our attempt to reduce traffic and not to scan IPv4 but to collect the services via domain names). The rest of the ASes agree with our measurement results.

%\vspace{-10pt}
{\bf Agents Active Measurements.} Agents with active probes found 608 ASes that were found not to be enforcing ingress filtering using the agents approach of the Spoofer Project (these include duplicates with the traceroute loops measurements). Those contain some of the duplicates from traceroute measurements: together both approaches of the Spoofer Project found 1,113 ASes to be spoofable. Apart from 57 ASes not included in our tests, we could not test 9 ASes, the rest were also identified by our tests.

Although the agents provide the optimal setup for testing filtering, with control over the packets that can be crafted and sent from both sides, as we explain in Related Work Section \ref{sc:works}, this approach is limited only to networks that deploy agents on their networks. In contrast, SMap provides better coverage since it is potentially applicable to every network that has one of the services that are required in our tests.

In total, our results identified 51,046 ASes to be spoofable, which is more than 80\% of the ASes that we tested. This is also 50,023 ASes more than that both the traceroute and the agents approaches found.

These findings show that SMap offers benefits over the existing methods, providing better coverage of the ASes in the Internet and not requiring agents or conditions for obtaining traceroute loops, hence improving visibility of networks not enforcing ingress filtering.

%% file: net-analysis.tex
\section{Networks Analysis}\label{sc:nets}
In order to understand if there are differences in enforcement of ingress filtering between different network types and different countries, we perform characterisation of the networks that we found to not be filtering spoofed packets. Specifically, we ask the following questions: {\em Does business type of networks or geo-location of networks influence filtering of spoofed packets?}

To derive the geo-location of ASes we used MaxMind GeoLite2 GeoIP database \cite{geolite2}. The results are listed in Table \ref{tab:as-by-country}. The tested ASes are distributed across different countries, with most ASes being in large countries, like US and Russia. The ration of spoofable ASes ranges between 67\% and 84\%, with Ukraine leading with the fraction of spoofable networks, with 84\%.
Surprisingly the ratio between the geolocation and spoofed packets is similar across different countries, with USA and Russia leading with 32\% of the networks and 33\% of the networks respectively, that do not filter spoofed packets. %; see Table \ref{tab:as-by-country}. 

\begin{table}[hbt!]
%\vspace{-10pt}
\centering
\begin{tabular}{ |p{1.5cm}|p{1.5cm}|p{1.5cm}|p{1.5cm}| }
 \hline
  Country & Tested ASes & Spoofable ASes & Spoofable Ratio \\
 \hline
  US	& 16,138	& 12,385	& 76.74\% \\ 
  \hline
  BR	& 7,692	& 6,447	& 83.81\% \\ 
  \hline
  RU	& 4,906	& 4,221	& 86.04\% \\ 
  \hline
  PL	& 2,092	& 1,739	& 83.13\% \\ 
  \hline
  DE	& 2,171	& 1,677	& 77.25\% \\ 
  \hline
  GB	& 2,231	& 1,648	& 73.87\% \\ 
  \hline
  UA	& 1,776	& 1,547	& 87.11\% \\ 
  \hline
  IN	& 1,970	& 1,480	& 75.13\% \\ 
  \hline
  ID	& 1,412	& 1,236	& 87.54\% \\ 
  \hline
  AU	& 1,625	& 1,234	& 75.94\% \\ 
  \hline
  CA	& 1,484	& 1,184	& 79.78\% \\ 
  \hline
  FR	& 1,310	& 1,036	& 79.08\% \\ 
  \hline
  NL	& 1,308	& 1,026	& 78.44\% \\ 
  \hline
  IT	& 1,013	& 850	& 83.91\% \\ 
  \hline
  ES	& 1,001	& 783	& 78.22\% \\ 
  \hline
  AR	& 918	& 733	& 79.85\% \\ 
  \hline
  RO	& 962	& 720	& 74.84\% \\ 
  \hline
  JP	& 782	& 606	& 77.49\% \\ 
  \hline
  HK	& 743	& 565	& 76.04\% \\ 
  \hline
  CZ	& 673	& 560	& 83.21\% \\ 
  \hline
\end{tabular}
\caption{Top-20 Countries with most tested ASes.}
%\vspace{-10pt}
\label{tab:as-by-country}
\end{table}

\begin{figure} [t!]
  \begin{center}
  %\vspace{-10pt}
   \includegraphics[width=0.45\textwidth]{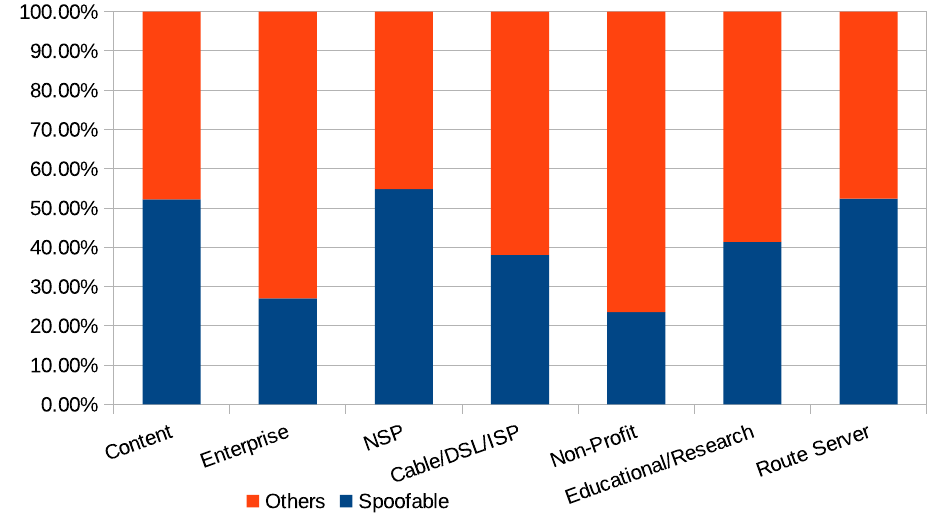}
   	%\vspace{-10pt}
		\caption{Spoofable ratio across ASes' types. AS type is queried from PeeringDB.}
		\label{fig:net_type_spoofable}
		%\vspace{-10pt}
  \end{center}
\end{figure}

 We also want to understand the types of networks that we could test via domains-wide scans. To derive the business types we use the PeeringDB. We classify the ASes according to the following business types: content, enterprise, Network Service Provider (NSP), Cable/DSL/ISP, non-profit, educational/research, route server at Internet Exchange Point (IXP)\footnote{A route server directs traffic among Border Gateway Protocol (BGP) routers.} We plot the networks that do not enforce ingress filtering according to business types in Figure \ref{fig:net_type_spoofable}. According to our study enterprise and non-profit networks enforce ingress filtering more than other networks. In contrast, NSPs contain the most networks that do not enforce ingress filtering.

\begin{figure} [t!]
  \begin{center}
  %\vspace{-15pt}
   \includegraphics[width=0.45\textwidth]{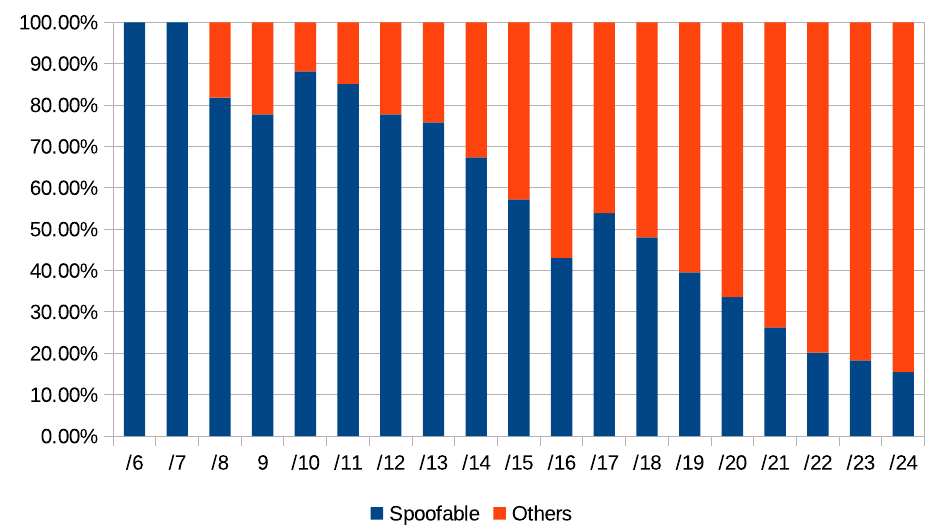}
   %\vspace{-10pt}
		\caption{Spoofable ratio according to networks' sizes. Network size is calculated from GeoLite2-ASN database.}
	%	\vspace{-10pt}
		\label{fig:net_size_spoofable}
		%\vspace{-10pt}
  \end{center}
\end{figure}

\begin{figure} [ht]
  \begin{center}
  %\vspace{-5pt}
   \includegraphics[width=0.48\textwidth]{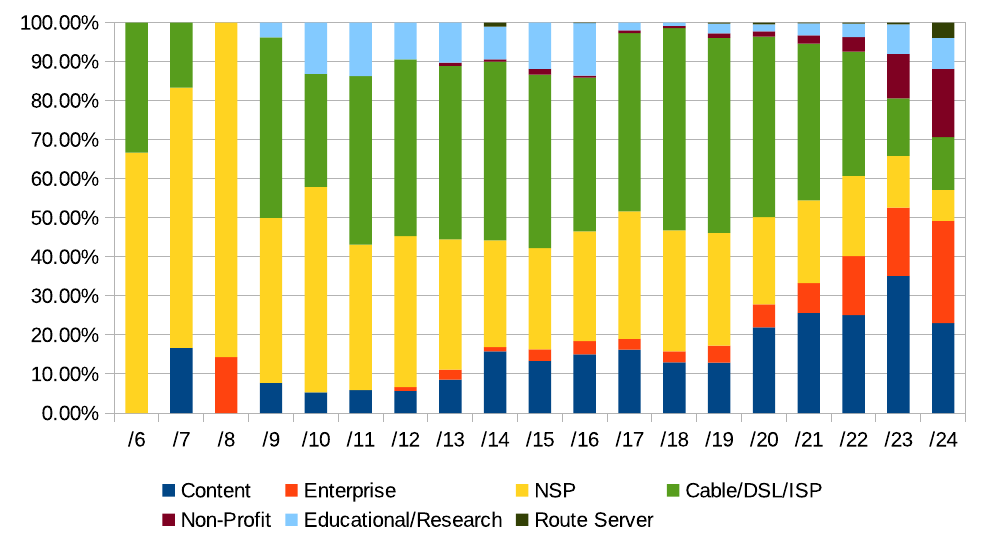}
   %\vspace{-15pt}
		\caption{Distribution of networks' sizes vs types.}
		\label{fig:net_type_size}
		%\vspace{-10pt}
  \end{center}
\end{figure}

%\vspace{-10pt}
There is a strong correlation between the AS size and the enforcement of spoofing, see Figure \ref{fig:net_size_spoofable}. Essentially, the larger the AS, the higher the probability that our tools identify that it does not filter spoofed packets. The reason can be directly related to our methodologies and the design of our study: the larger the network the more services it hosts. This means that we have more possibilities to test if spoofing is possible: for instance, we can identify a higher fraction of servers with a globally incremental IPID counters, which are not ``load balanced''. In Figure \ref{fig:net_type_size} we plot the statistics of the tested networks according to their size and type. The results show a correlation between the size of the network and its type. For instance, most NSP networks are large, with CIDR/6. This is aligned with our finding that among NSP networks there was the highest number of spoofable networks. %In Appendix, we plot the fraction of spoofable networks according to networks' sizes for each network type separately, i.e., for DSL/ISP in Appendix, Figure \ref{fig:net_size_isp}, for NSP in Appendix, Figure \ref{fig:net_size_nsp}, for enterprises in Appendix, Figure \ref{fig:net_size_ent}, for content providers in Appendix, Figure \ref{fig:net_size_content}, for research organisations in Appendix, Figure \ref{fig:net_size_research}, for non-profit organisations in Appendix, Figure \ref{fig:net_size_np}, and for route server networks in Appendix, Figure \ref{fig:net_size_route}.

%%%%%%%%%%%%%%%%%%%%%%%%%%%%%%%

\ignore{
\section{Tracking Filtering Adoption}
Researchers have been trying to understand the adoption of ingress filtering and through active measurements, \cite{mauch2013open, kuhrer2014exit}. In these cases the measurements were either performed for one time studies or have been (and are being) performed on specific networks, such as those operating agents by volunteers or by paid workers. In any case, all of the proposlas do not enable performing comprehensive scans of the Internet networks. SMap lowers the barriers to performing Internet-scale studies
}
% Which studies have recurring measurements.
% which are one time studies (like kuhrer or paid workers).

%% file: conclusions.tex
\section{Conclusions}\label{sc:conc}
Much effort is invested to understand the extent of spoofability in the Internet. However, current measurement studies have limited applicability, providing results that apply to a small set of Internet networks. 

Our work provides the first comprehensive view of ingress filtering in the Internet. We showed how to improve the coverage of the Internet in ingress filtering measurements to include many more ASes that were previously not studied. Our techniques allow to cover more than 90\% of the Internet ASes, in contrast to best coverage so far of 7.5\% of the ASes performed by the Spoofer Project. This coverage can be further extended to include 100\% of the Internet's ASes by scanning the IPv4 range instead of opting for the dataset of \cite{sonar}, that we used in our study. %In this work our goal was to explore: given the minimal fraction of traffic generated, what is the lower bound on coverage and accuracy that can be achieved. In DNS looup and PMTUD tests the traffic volume that we generate is similar  to that of the Spoofer Project (which generates the least traffic in comparison to the other active measurements). The traffic that IPID technique requires is proportional to the traffic volume that the server in the tested network receives. Nevertheless, even with the minimal traffic volume that we produce, the effectiveness of SMap is higher than those of previous studies: a comparative analysis of our results with the datasets published by the other tools showed that we found new ASes which were not previously identified as not filtering spoofed traffic. SMap also uses a much more stable infrastructure than the other studies which enables running more accurate longitudinal studies: for instance, since July 2019 less than 1\% of periodically tested ASes became N/A.

The most significant aspect of our methodologies is that they do not require coordination with the scanned networks. SMap can measure spoofability in any TCP/IP network with standard and widely supported services, such as Email and web. We integrated into SMap three techniques for testing ingress filtering: DNS-based, IPID-based and PMTUD-based. Our experimental comparison of the effectiveness of the techniques demonstrated that DNS-based technique has a wider applicability rate on networks that operate DNS resolvers than the other two techniques, while the detection of the spoofability of networks is more accurate with PMTUD. %Our analysis shows that the measurements performed by SMap provide a lower bound on the networks that do not enforce ingress filtering. 

We set up SMap as a public service for continuous collection and analysis of the ingress filtering in the Internet at\\{\tt https://smap.cad.sit.fraunhofer.de}.

%As a future research direction we recommend extending SMap with more measurements techniques. %The design of SMap is modular to allow for easy integration of new measurement techniques.

%SMap can be accessed at We provide open source access to the code of SMap, as well as to the datasets collected by SMap and the statistics, at \url{https://smap.cad.sit.fraunhofer.de}. 

%Future research direction we propose to extend the applicability of the IPID technique. As an initial feasibility step in this work we focused on incremental IPID counters. As a next step, we suggest to apply the reverse engineering techniques from \cite{DBLP:journals/corr/abs-2012-07432,DBLP:journals/corr/abs-2008-12981} over the pseudorandom IPID counters. This would allow using IPID technique also with networks that have services with unpredictable IPID counters. 

%We provide access to SMap at \url{http://141.12.16.39/}, with publicly available datasets and statistics of the ingress filtering measurements in the Internet.

% tradeoff between traffic and accuracy of detection of spoofing.
% to explain why it can be that the other tools missed out on identifying the other ASes that do not filter.